\documentclass[prl,aps,a4paper,twocolumn,superscriptaddress,nofootinbib]{revtex4-1}

\usepackage[utf8x]{inputenc}
\usepackage{amssymb}
\usepackage{amsmath}
\usepackage{graphicx}
\usepackage{bbm}
\usepackage{psfrag}
\usepackage{latexsym}
\usepackage{hyperref}
\usepackage{color}

\newcommand{\be}{\begin{equation}}
\newcommand{\ee}{\end{equation}}
\newcommand{\bea}{\begin{eqnarray}}
\newcommand{\eea}{\end{eqnarray}}
\newcommand{\ba}{\begin{eqnarray}}
\newcommand{\ea}{\end{eqnarray}}

\newcommand{\beq}{\begin{equation}}
\newcommand{\eeq}{\end{equation}}
\newcommand{\beqa}{\begin{eqnarray}}
\newcommand{\eeqa}{\end{eqnarray}}
\newcommand{\beqar}{\begin{eqnarray*}}
\newcommand{\eeqar}{\end{eqnarray*}}

\newcommand{\eq}{\begin{equation}}
\newcommand{\eqx}{\end{equation}}
\newcommand{\eqn}{\begin{eqnarray}}
\newcommand{\eqnx}{\end{eqnarray}}

\newcommand{\ket}[1]{{\left| #1 \right\rangle}} 
\newcommand{\bra}[1]{{\left\langle #1 \right|}} 
\newcommand{\braket}[2]{{\left\langle #1 \mid #2 \right\rangle}} 

\newcommand{\Span}{{\mathsf{Span}}} 

\newcommand{\cP}{{\mathcal P}} 

\newcommand{\cG}{{\mathcal G}} 
\newcommand{\cC}{{\mathcal C}} 

\usepackage[normalem]{ulem}

\newcommand{\nobracket}{}

\begin{document}

\author{Shira Chapman}\email{schapman@perimeterinstitute.ca}

\affiliation{\it Perimeter Institute for Theoretical Physics, Waterloo, Ontario N2L 2Y5, Canada}

\author{Michal P. Heller}\email{michal.p.heller@aei.mpg.de}\homepage{aei.mpg.de/GQFI}

\altaffiliation[\\On leave from: ]{\emph{National Centre for Nuclear Research, 00-681 Warsaw, Poland.}}

\affiliation{Max Planck Institute for Gravitational Physics, Potsdam-Golm, D-14476, Germany}

\author{Hugo Marrochio}\email{hmarrochio@perimeterinstitute.ca}

\affiliation{\it Perimeter Institute for Theoretical Physics, Waterloo, Ontario N2L 2Y5, Canada}

\affiliation{\it Department of Physics $\&$ Astronomy, University of Waterloo, Waterloo, ON N2L 3G1, Canada}

\author{Fernando Pastawski}\email{fernando.pastawski@gmail.com}

\affiliation{\it \mbox{Dahlem Center for Complex Quantum Systems, Freie Universit{\"a}t Berlin, Berlin, D-14195, Germany}}
\affiliation{Max Planck Institute for Gravitational Physics, Potsdam-Golm, D-14476, Germany}

\title{Towards Complexity for Quantum Field Theory States}

\begin{abstract}
We investigate notions of complexity of states in continuous quantum-many body systems. We focus on Gaussian states which include ground states of free quantum field theories and their approximations encountered in the context of the continuous version of Multiscale Entanglement Renormalization Ansatz.
Our proposal for quantifying state complexity is based on the Fubini-Study metric.
It leads to counting the number of applications of each gate (infinitesimal generator) in the transformation, subject to a state-dependent metric.
We minimize the defined complexity with respect to momentum preserving quadratic generators which form $\mathfrak{su}(1,1)$ algebras.
On the manifold of Gaussian states generated by these operations the Fubini-Study metric factorizes into hyperbolic planes with minimal complexity circuits reducing to known geodesics.
Despite working with quantum field theories far outside the regime where Einstein gravity duals exist, we find striking similarities between our results and those of holographic complexity~proposals.
\end{abstract}

\maketitle

\noindent \emph{1. Introduction.--}
Applications of quantum information concepts to high energy physics and gravity have recently led to many far-reaching developments. In particular, it has become apparent that special properties of entanglement in holographic~\cite{Maldacena:1997re} quantum field theories (QFT)s states are crucial for the emergence of smooth higher-dimensional (bulk) geometries in the gauge-gravity duality~\cite{VanRaamsdonk:2010pw}. Much of the progress in this direction was achieved building on the holographic entanglement entropy proposal by Ryu and Takayanagi~\cite{Ryu:2006bv} which geometrizes the von Neumann entropy of a reduced density matrix of a QFT in a subregion in terms of the area of  \mbox{codimension-2} bulk minimal surfaces anchored at the boundary of this subregion (see e.g., Ref.~\cite{Rangamani:2016dms} for a recent overview).  However, Ryu-Takayanagi surfaces are often unable to access the whole holographic geometry~\cite{Balasubramanian:2014sra,Susskind:2014moa,Freivogel:2014lja}.
This observation has led to significant interest in novel, from the point of view of quantum gravity, \mbox{codimension-1} (volume) and \mbox{codimension-0} (action) bulk quantities, whose behavior suggests conjecturing a link with the information theory notion of quantum state complexity~\cite{Susskind:2014rva,Stanford:2014jda,Alishahiha:2015rta,Brown:2015bva,Brown:2015lvg,Carmi:2016wjl,Chapman:2016hwi}. 
In fact, a certain identification between complexity and action was originally suggested by Toffoli \cite{Toffoli1998,Toffoli1999} outside the context of holography.

Quantum state complexity originates from the field of quantum computations, which are usually modeled in a finite Hilbert space as the application of a sequence of gates chosen from a discrete set. In this context, the complexity of a unitary $U$ is roughly associated with the minimal number of gates necessary to realize (or approximate)~$U$. Notable progress has been made in connecting this notion to distances in Riemannian geometries derived from a set of generators \cite{Nielsen2006}.
The complexity of a target state $\ket{T}$ is usually subordinated to unitary complexity by specifying a ``simple'' reference state $\ket{R}$ and minimizing the complexity of $U$ subject to $\ket{T}=U\ket{R}$~\cite{WatrousComp, Aaronson:2016vto};
Our approach differs in defining state complexity directly.

In the context of holography, the organization of discrete tensor networks (seen as a quantum circuit $U$) has been suggested to give a qualitative picture of how quantum states give rise to emergent geometries~\cite{Swingle2012}.
This heuristic analysis was applied to the multiscale entanglement renormalization ansatz (MERA)~\cite{MERA}, employed to find ground states of critical physical theories presenting a tensor network structure reminiscent of an AdS time slice.
This motivated proposing ``complexity equals volume'' (CV)~\cite{Stanford:2014jda} and ``complexity equals action'' (CA) \cite{Brown:2015bva,Brown:2015lvg} as new entries in the holographic dictionary.
However, in holography one naturally considers continuum setups, QFTs, and there are shortcomings of traditional approaches to complexity when attempting to address field theory states.
The aim of this letter is to bridge a pressing gap by exploring complexity-motivated distance measures in QFTs.

The main challenges in providing a workable definition of complexity in the continuum are related to  choosing: a)~a~reference state $\ket{R}$,
b)~a~set of allowed gates (correspondingly infinitesimal generators),
c)~a measure for how such gates contribute to the resulting distance function  and a procedure for how to minimize it, d)~a way to regulate ultraviolet (UV) divergences.
Our proposed choice for c) is to measure the path length
 by integrating the Fubini-Study (FS) line element along a path from $\ket{R}$ to $\ket{T}$ associated to an allowed realization of $U$.
Minimizing the path will amount to studying geodesics on the manifold of quantum states induced by allowed gates acting on the reference state.
In this way, our approach derives complexity from the projective structure of the Hilbert space in a universal way.
In the FS prescription, directions which modify the state by an overall phase have no effect on the complexity. Simultaneously with our work, Ref.~\cite{Jefferson:2017sdb} appeared which considers a different approach based on unitary complexity \cite{Nielsen2006} (see \cite{SupMat}F for a comparison).

While the FS prescription is quite general, our choices for a), b) and d) render the necessary calculations tractable.
Some of these choices, are inspired by the continuous MERA (cMERA) approach to free QFTs \cite{Haegeman:2011uy,Nozaki:2012zj,Mollabashi:2013lya},
which we briefly review in \cite{SupMat}A.
Similarly to the states in cMERA, our choices for the reference state $\ket{R}$ and target state $\ket{T}$ will be {\it pure Gaussian states} and allowed generators will be subsets of quadratic operators.
Our choices include cMERA in the set of allowed circuits, letting us test  its optimality.
We perform our analysis in momentum space and ignore  frequencies above the UV cutoff $\Lambda$ which equips  us with a notion of approximation.
Unlike in cMERA, $\Lambda$ need not coincide with the reference state characteristic scale $M$, defined below in Sec.~2, since the freedom of choosing the reference state is a part of the definition of  complexity and is a priori independent from a notion of cutoff or regulator
 (this observation is due to R.~C.~Myers).

As a first step, we consider the two mode squeezing operator for each pair of opposite momenta $\pm\vec{k}$.
We then extend our analysis to include the full set of momentum preserving quadratic generators which form $\mathfrak{su}(1,1)$ algebras.
In this case the study of minimal complexity reduces to the study of geodesics on a product of hyperbolic planes.

While a full literature review is outside the scope of this article, there is a substantial body of important recent developments which include e.g., Refs.~\cite{Miyaji2015,Molina-Vilaplana2015,Caputa:2017urj,Caputa:2017yrh, Czech2017,Brown:2017jil,Hashimoto:2017fga,Roberts2017,Chemissany:2016qqq}.

\vspace{6 pt}

\noindent \emph{2. Complexity from the Fubini-Study metric.--}
We are interested in considering unitary operators $U$ arising from iterating  generators $G(s)$ taken from some elementary set of Hermitian operators $\cG$.
The allowed transformations $U$ can then be represented as  path ordered exponentials
\begin{align}\label{eq.defU}
U(\sigma) = \cP e^{- i \int_{s_{i}}^{\sigma}  G(s) ds}\, .
\end{align}
Here, $s$ parametrizes progress along a path, starting at $s_{i}$ and ending at $s_f$ and $\sigma \in [s_i,s_f]$ is some intermediate value of $s$.
The path-ordering $\cP$ is required for non-commuting generators $G(s)$.
We seek a path achieving
$\ket{T} \approx U(s_{f}) \ket{R}$, where $(\approx)$ indicates that states coincide for momenta below a cutoff $\Lambda$.
According to the FS line element (see e.g., \cite{bengtsson2007geometry}),
\begin{align}
\label{eq.FSmetric}
ds_{FS}(\sigma) = d\sigma \sqrt{
\Big|\partial_{\sigma} \ket{\Psi(\sigma)}\Big|^{2} - \Big| \bra{\Psi(\sigma)}\,  \partial_{\sigma} \ket{\Psi(\sigma)}\Big|^{2}
}\, ,
\end{align}
the length of a path going via states $\ket{\Psi(\sigma)}$ is
\begin{align}\label{eq.length}
\ell(\ket{\Psi(\sigma)}) = \int_{s_{i}}^{s_{f}} ds_{FS}(\sigma)\, .
\end{align}
For a path $\ket{\Psi(\sigma)} = U(\sigma) \ket{R}$, with $U(\sigma)$  given by Eq.~\eqref{eq.defU}, the line element of Eq.~\eqref{eq.length} can be re-expressed as
\begin{align}
\label{eq.FSvariance}
ds_{FS}(\sigma) = d\sigma \sqrt{\langle G^2(\sigma)\rangle_{\Psi(\sigma)} - \langle G(\sigma)\rangle_{\Psi(\sigma)}^2
}\, ,
\end{align}
and is independent of path reparametrizations.

If the path $\ket{\Psi(\sigma)}$ is unrestricted, the unique unitarily invariant distance measure $d_{R,T}=\arccos\left| \braket{R}{T}\right| \leq \pi/2$ is obtained.
However, by restricting the allowed generators $G(s)$,  highly non-trivial notions of distance deserving the name complexity may be obtained.
Our proposal is to define the complexity ${\cal C}$ as the minimal length according to Eq.~\eqref{eq.length} of a path from $\ket{\Psi(s_i)}\approx \ket{R}$ to $\ket{\Psi(s_f)}\approx \ket{T}$ driven by generators $G(s)$ in $\cG$
\begin{align}
\label{eq.Cdef}
\cC(\ket{R},\ket{T},\cG, \Lambda) = \min_{G(s)} \ell(\ket{\Psi(\sigma)})\,.
\end{align}
The proposed complexity $\cC$ inherits the properties of a distance function from the FS metric.

\vspace{10 pt}

\noindent \emph{3. Gaussian states in free QFTs.-- }
We consider a theory of free relativistic bosons in $(d+1)$-spacetime dimensions defined by the quadratic Hamiltonian
\begin{align}
\label{eq.hamiltonian}
H_m = \int  d^{d}x : \left\{ \frac{1}{2} \pi^{2} + \frac{1}{2} (\partial_{\vec{x}} \phi)^{2} + \frac{1}{2} m^{2} \phi^{2} \right\}:
\end{align}
with commutation relations \mbox{$[\phi(\vec{x}), \pi(\vec{x}')] = i \, \delta^{d}(\vec{x} - \vec{x}')$}.
This theory describes noninteracting particles created and annihilated by operators $a_{\vec{k}}^{\dagger}$ and $a_{\vec{k}}$ obeying $[a_{\vec{k}}, a_{\vec{k}'}^{\dagger}] = \delta^{d} (\vec{k} - \vec{k}')$.
These are related to the field and momentum operators via ($\omega_k \equiv \sqrt{k^2+m^2}$)
\small
\begin{align}
\hspace{-5 pt}\phi(\vec{k}) = \frac{1}{\sqrt{2 \omega_k}} \left( a_{\vec{k}}+a_{-\vec{k}}^\dagger \right) \,\, \mathrm{and} \,\,
\pi(\vec{k}) = \frac{\sqrt{\omega_k}}{\sqrt{2} i} \left(a_{\vec k}-a_{-{\vec k}}^\dagger\right)
\end{align}
\normalsize
and diagonalize the Hamiltonian: $H_m = \int d^{d}k \, \omega_{k} \, a_{\vec{k}}^{\dagger}\, a_{\vec{k}}$.
For $m = 0$ we obtain a free \textit{conformal field theory} (CFT).

A general translation invariant {\it pure Gaussian state} $\ket{S}$
with momentum space correlation functions
\begin{align}
\label{eq.corr.alpha.k}
\bra{ S} \phi(\vec{k}) \phi(\vec{k}') \ket{ S }= \frac{1}{2 \, \alpha_{k}} \delta^{(d)} \left( \vec{k} + \vec{k}' \right),
\end{align}
is specified by its nullifiers (annihilation operators):
\begin{align}\label{eq.vacuumNullifiers}
\left\{\sqrt{\frac{\alpha_{k}}{2}} \phi(\vec{k}) + i \, \frac{1}{\sqrt{2 \, \alpha_{k}}} \pi(\vec{k}) \right\} \ket{S} = 0\,.
\end{align}

The ground state $\ket{m}$ of the free theory \eqref{eq.hamiltonian} is a pure Gaussian state corresponding to $\alpha_k = \omega_k$.
The ground state $\ket{m}$ is a product of vacuum states in momentum space without particles according to the number operators $n_{\vec{k}}\equiv a^\dagger_{\vec{k}} a_{\vec{k}} $.
In momentum space, the only nontrivial correlations in $\ket{S}$ are between $\vec{k}$ and~$(-\vec{k})$ modes.
In real-space, the $\vec{k}$-dependent factor on the RHS of Eq.~\eqref{eq.corr.alpha.k} leads to  spatial correlations (and entanglement).

A natural choice for a reference state $\ket{R(M)}$ is the Gaussian state corresponding to
\begin{equation}\label{eq.ReferenceState}
\ket{R(M)}: \qquad \alpha_k=M\, .
\end{equation}
Since here $\alpha_k$ is independent of $k$ this state is a product state with no spatial correlations, i.e., in real space the two point function of field operators takes the form $\bra{ R(M)} \phi(\vec{x}) \phi(\vec{x}') \ket{R(M)} = \frac{1}{2 M}\delta^{d}(\vec{x} - \vec{x}')$. Nevertheless in the basis associated with energy eigenstates of $H_{m}$ momentum sectors $\vec{k}$ and $-\vec{k}$ are~pairwise~entangled according to \eqref{eq.corr.alpha.k}.
We will show that the reference state scale $M$ is related to certain ambiguities encountered in the context of holographic complexity.
The annihilation and creation operators $b_{\vec k}$ and $b^\dagger_{\vec k}$ associated with the state $\ket{R(M)}$
can be related to those of the vacuum state $\ket{m}$ by the following Bogoliubov transformation
\begin{equation}\label{eq.bopts}
\begin{split}
&~~~b_{\vec{k}} = \beta^+_{k} a_{\vec{k}} + \beta^-_{k} a^\dagger_{-\vec{k}}; \qquad b_{\vec{k}}|R(M)\rangle=0;
\\
&\beta^+_k = \cosh{2r_k};
 ~~~\beta^-_k = \sinh{2r_k}; ~~~ r_k \equiv \log\sqrt[4]{\frac{M}{\omega_{k}}}\,.
\end{split}
\end{equation}

As our target state, we consider the approximate ground state $\ket{m^{(\Lambda)} }$ characterized by the  UV momentum cut-off $\Lambda$ which corresponds to:
\begin{align}
\label{eq:alpha}
\ket{m^{(\Lambda)} }: ~~~\alpha_k = \left\{ \begin{array}{cl}
  \omega_{k},&  ~~k < \Lambda~~\mbox{(QFT vacuum)}
  \\
  M,& ~~ k \geq \Lambda~~\mbox{(product state)}
  \end{array}\right.,
\end{align}
with correlation functions interpolating between those of the vacuum state $\ket{m}$ and the reference state $\ket{R(M)}$  as momentum increases according to Eq.~\eqref{eq.corr.alpha.k}. This state is in fact identical to the real ground state $\ket{m}$ up to the cut-off momentum.
When $M=\omega_\Lambda$, this state is identical to the one obtained by cMERA circuits \cite{Haegeman:2011uy,Nozaki:2012zj} (see e.g. Ref.~\cite{Hu:2017rsp}).

The target states \eqref{eq:alpha} can be reached from the reference states  \eqref{eq.ReferenceState} by a circuit constructed with \textit{two mode squeezing operators} which entangle the $\vec{k}$ and $-\vec{k}$ modes,
\small
\begin{align}
\label{eq.defK}
\begin{split}
\hspace{-6 pt}K(\vec{k}) &= \phi(\vec{k}) \pi(-\vec{k}) + \pi(\vec{k}) \phi(-\vec{k}) \\
&= i \left(a_{\vec{k}}^{\dagger} \, a_{-\vec{k}}^{\dagger} - a_{\vec{k}} \, a_{-\vec{k}} \right) = i \left(b_{\vec{k}}^{\dagger} \, b_{-\vec{k}}^{\dagger} - b_{\vec{k}} \, b_{-\vec{k}} \right).
\end{split}
\end{align}
\normalsize
This operator is the main building block in cMERA circuits, and allows
preparing the target state as follows
\begin{align}
\label{eq.cMERAcontractGEN}
\ket{ m^{(\Lambda)} } = e^{- i \int_{k \leq \Lambda} d^{d} k \,  r_k \,  K(\vec{k})  } \, \ket{ R(M) }\, ,
\end{align}
which is the starting point for our complexity analysis.

\vspace{10 pt}

\noindent \emph{4. Ground state complexity with a single generator per pair of momenta $\pm\vec{k}$.-- }
We start by evaluating our proposed complexity under the assumption that we allow for a single generator per pair of momenta $\pm\vec{k}$ which we take to be $K(\vec k)$ of Eq.~\eqref{eq.defK}, i.e., $\cG = \Span\{K( \vec{k})\}$, where $\Span$ is taken over the field of real numbers.
These generators continue to achieve minimal complexity within the larger $\mathfrak{su}(1,1)$ class considered in Sec.~5.
We consider circuits of the form \eqref{eq.defU} with
\begin{align}\label{eq.defG}
G(\sigma) = \int_{k \leq \Lambda} d^{d} k\,  K(\vec{k}) y_{\vec{k}}(\sigma)\, .
\end{align}
Since all the $K(\vec{k})$ commute, the unitary $U(\sigma)$ of \eqref{eq.defU} is simply specified by the integrated values
\begin{equation}\label{eq.DeltaY}
Y_{\vec{k}}(\sigma) := \int_{s_i}^{\sigma} y_{\vec{k}}(s)ds; \qquad Y_{\vec{k}}(s_{f})  = r_k,
\end{equation}
where $Y_{\vec{k}}(s_{f})$ was fixed to match Eq.~\eqref{eq.cMERAcontractGEN}.
The commutation of generators allows the variance in the FS line element \eqref{eq.FSvariance} to be evaluated at any state $\ket{\Psi(\sigma)}$ along the path.
Furthermore, the variance is additive with respect to the different $K(\vec{k})$ contributions because only equal or opposite momenta can be correlated.
The complexity minimization of Eq.~\eqref{eq.Cdef} then reduces to
\begin{align}
\label{eq.C2}
{\cal C} = \min_{Y_{\vec{k}}(\sigma)} \int_{s_i}^{s_{f}} d\sigma \, \sqrt{
 2\, \mathrm{Vol} \, \int_{k \leq \Lambda} d^{d} k  \,
	\left( \partial_\sigma Y_{\vec{k}}(\sigma) \right)^{2}},
\end{align}
where $\mathrm{Vol} \equiv \delta^d(0)$ is the volume of the $d$-dimensional time slice. One recognizes a flat Euclidean geometry with coordinates $Y_{\vec{k}}(\sigma)$ continuously labeled by $\vec{k}$.
To achieve minimal complexity the generators for the different momenta must act simultaneously with ratio dictated by Eq.~\eqref{eq.DeltaY} (straight path).
A particularly simple affine parametrization for the path is
\small
\begin{equation}\label{eq.TheCircuit}
Y_{\vec{k}}(\sigma) = \frac{\sigma - s_{i}}{s_{f} - s_{i}} \,Y_{\vec{k}}(s_f);
~~~ y_{k}(\sigma) = \frac{1}{s_{f} - s_{i}} \,Y_{\vec{k}}(s_f).
\end{equation}
\normalsize
As the corresponding cMERA circuit presents a $\sigma$ dependent ratio, the complexity associated with it will generically be larger (as shown in \cite{SupMat}A).
Evaluating \eqref{eq.C2} with \eqref{eq.TheCircuit}, the minimal complexity reads
\begin{align}\label{eq.C2explicit}
{\cal C}^{(2)} = \sqrt{ 2 \, \mathrm{Vol} \, \int_{k \leq \Lambda} d^{d} k  \,  r_k ^{2}}\, ,
\end{align}
where the superscript $(2)$ anticipates an interpretation of Eq.~\eqref{eq.C2explicit} as an $L^{2}$ norm.

Suppose on the other hand that $\cG$ contains only individual generators $K(\vec{k})$ and not their linear span.
This is analogous to disallowing different elementary gates in a circuit to act simultaneously.
Our path parameters in this case consist of $\sigma$ and $\vec k$.
The arguments leading to Eq.~\eqref{eq.C2} continue to hold except that now, the $k$ integral must be pulled out of the square root and an extra $\sqrt{\mathrm{Vol}/2}$ factor appears.
This leads to an $L^{1}$ norm (Manhattan distance) complexity
\begin{align}\label{eq.C1NormNN}
{\cal C}^{(1)} = \,\mathrm{Vol} \, \int_{k \leq \Lambda} d^{d} k  \, \left| r_k \right|\, .
\end{align}
More generally, and without reference to the FS metric, one can postulate  $L^{n}$ norms as a measure of complexity
\begin{equation}
\label{eq.complexityproposal}
{\cal C}^{(n)} = 2 \sqrt[n]{\frac{\mathrm{Vol}}{2} \, \int_{k \leq \Lambda} d^{d} k  \, \left| r_k \right|^{n}}.
\end{equation}
\normalsize
The leading divergence in the complexity measures ${\cal C}^{(n)}$ is proportional to
\begin{equation}\label{eq.LnNormsDivB}
{\cal C}^{(n)} \sim
\mathrm{Vol}^{1/n}  \Lambda^{d/n}\log (M/\Lambda),
\end{equation}
when $M$ and $\Lambda$ are chosen independently and to
\begin{equation}\label{eq.LnNormsDivA}
{\cal C}^{(n)} \sim  \mathrm{Vol}^{1/n}  \Lambda^{d/n}
\end{equation}
when $M=\Lambda$.
See  \cite{SupMat}B for some additional details on evaluating the ground state complexities using the $\mathcal{C}^{(n)}$ measures.
The ${\cal C}^{(1)}$ norm results carry resemblance to those found in the context of holographic complexity as we explain in Sec.~6.

\vspace{10 pt}

\noindent \emph{5. Ground state complexity using $\mathfrak{su}(1,1)$ generators.-- }
Here, we extend our minimization to a larger set of generators $\cG$ that transforms $\ket{R(M)}$ into $\ket{m^{(\Lambda)}}$.
Namely, we consider momentum preserving quadratic operators, which for each  $\vec{k}$ are spanned by
\small
\begin{align}\label{eq.SU(1,1)a}
\begin{split}
&\cG = \Span\left\{ K_0, K_1 \equiv\frac{K_+ + K_-}{2} , K_2 \equiv \frac{K_+ - K_-}{2i}\right\} \\
&	K_+ =\frac{b^\dagger_{\vec{k}}b^\dagger_{-\vec{k}}}{2},~~~
	K_-=\frac{b_{\vec{k}}b_{-\vec{k}}}{2},~~~
	K_0=\frac{b^\dagger_{\vec k} b_{\vec k} +  b_{-\vec k} b^\dagger_{-\vec k}  }{4}.
\end{split}
\end{align}
\normalsize
These Hermitian operators form a larger (yet manageable), algebraically closed extension of the generators $K = -4K_2$ of Eq.~\eqref{eq.defK} used in cMERA circuits.
The algebra formed is an infinite product of $\mathfrak{su}(1,1)$ subalgebras of quadratic generators commuting with $n_{\vec{k}} - n_{-\vec{k}}$.
The path in Eqs.~\eqref{eq.defU}, \eqref{eq.defG} and \eqref{eq.TheCircuit} is contained in this larger set.
We prove that it continues to be minimal and determine its complexity, although we emphasize that this does not follow automatically from the results of the previous section.
For instance, in \cite{SupMat}D we study another constant generator $B(\vec k,M)$ which belongs to the extended
$\mathfrak{su}(1,1)$ subalgebras but does not lead to a minimal length path.
This generator has bounded norm and drives constant period oscillations between the reference state $\ket{R(M)}$ and target state $\ket{m^{(\Lambda)}}$.

We will see that the manifold of states generated by each $\mathfrak{su}(1,1)$ is a hyperbolic plane, one for each pair of opposite momenta.
Minimal complexity paths correspond to geodesics in the resulting tensor product manifold.
At the level of the state $| \Psi (\sigma) \rangle$, the most general $\mathfrak{su}(1,1)$ path can always be recast in the form (see \cite{SupMat}C)
\begin{equation}\label{path2}
 \hspace{-12 pt} | \Psi (\sigma) \rangle =  \mathcal{N}(\sigma) e^{\int d^d k \gamma_{+} ( \vec k,\sigma) K_{+}(\vec k)}| R (M) \rangle\,,
\end{equation}
where $\mathcal{N}(\sigma)$ is a complex normalization and $\sigma$ is the path parameter from Eq.~\eqref{eq.defU}.
This implies that the state $| \Psi (\sigma) \rangle$ can be conveniently parametrized by a single complex parameter $\gamma_+(\sigma)$.
The existence of spurious parameters is a manifestation of the non-uniqueness of the unitary circuit $U(\sigma)$ achieving a minimal complexity path.
Due to the non-commutative nature of the generators \eqref{eq.SU(1,1)a}, there is a non-trivial relationship between their coefficients in the path ordered exponential in Eq.~\eqref{eq.defU} and $\gamma_+(\vec{k},\sigma)$.
Our reference state $|R(M)\rangle$ corresponds to $\gamma_+(\vec{k},s_i)=0$ while the target state $\ket{m^{(\Lambda)}}$ corresponds to $\gamma_+(\vec{k},s_f)=\tanh (2 r_k)$.

\begin{figure}[ht]
	\centering
	\includegraphics[scale=0.058]{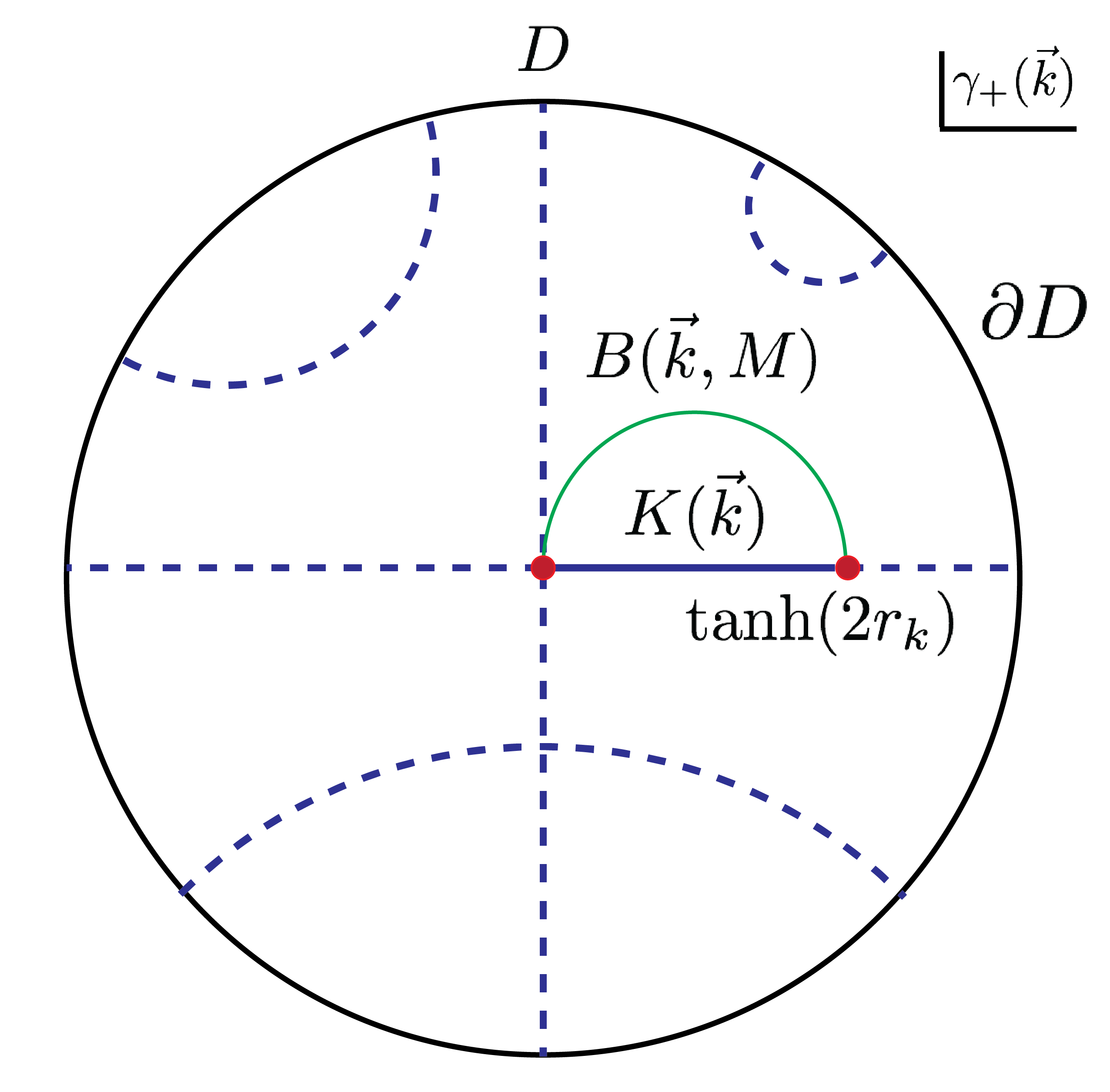}
	\caption{
		The Poincar\'e disk, parametrized by real (horizontal) and imaginary (vertical) components of $\gamma_+$. Examples of geodesics appear as  dashed lines.
		The two dots indicate the reference state  (the center) and the target state (on the real axis).
		The geodesic connecting the two is the straight solid line along the diameter, corresponding to the generator $K(\vec{k})$.
		The solid semicircle is the non-geodesic path generated by $B(\vec{k}, M)$ (see \cite{SupMat}D).
		The $\mathfrak{su}(1,1)$ algebra generates isometries on the hyperbolic plane. 		
		}
	\label{PoincareDisk}
\end{figure}

Evaluating the FS line element \eqref{eq.FSmetric} along the path \eqref{path2} leads to the following remarkably simple form
\begin{equation}\label{LineElemFS:su11}
ds_{FS}(\sigma) =  d \sigma \sqrt{ \frac{\text{Vol}}{2} \int_{\Lambda} d^d \vec{k} \frac{ \gamma'_{+} (\vec{k}, \sigma)  \gamma'^{\ast}_{+} (\vec{k}, \sigma) }{(1- |\gamma_{+}(\vec{k}, \sigma)|^2  )^2} },
\end{equation}
(see \cite{SupMat}C for the derivation).
This line element corresponds to a direct product of Poincar\'e disks parametrized by the complex coordinates  $\gamma_+(\vec{k})=\gamma_+(-\vec{k})$ ($|\gamma_+(\vec{k})|<1$), one for each pair of momenta $\pm\vec{k}$ (an example of such a disk is illustrated in Fig.~\ref{PoincareDisk}).
The Poincar\'e disk is the manifold naturally associated with the coset $SU(1,1)/U(1)$  (see e.g., \cite{Perelomov:1971bd,Provost:1980nc,Perelomov:1986tf}) and its structure of geodesics is well known.
Given an affinely parametrized geodesic on a Riemannian product manifold such as \eqref{LineElemFS:su11}, its natural projections are affinely parametrized geodesics within each factor manifold.
The relative speeds of these projections are coupled and will, as in \eqref{eq.TheCircuit}, be fixed by the target state.

The geodesic connecting $|R(M)\rangle$ and $|m^{(\Lambda)}\rangle$ follows the radial direction on the Poincar\'e disc which corresponds to the affinely parametrized path $\gamma_+(\vec k,\sigma) = \tanh(2  r_k \sigma)$, $\sigma\in[0,1]$, generated by $K(\vec k)$  of Sec.~4.
Therefore the path in Eqs.~\eqref{eq.defG}, \eqref{eq.TheCircuit} leads to minimal complexity even within the larger class of $\mathfrak{su}(1,1)$ generators.

\vspace{6 pt}

\noindent \emph{6. Comparison with Holographic Complexity Proposals.--}
There are two proposals for the gravity dual of  complexity in terms of maximal codimension-$1$ volumes  (CV~\cite{Stanford:2014jda}) or on-shell actions of the Wheeler-DeWitt patch bounded by null hypersurfaces (CA~\cite{Brown:2015bva,Brown:2015lvg}) in the dual bulk spacetime.  The structure of the vacuum UV divergences of holographic complexity can be characterized by a UV regularization scheme \cite{Carmi:2016wjl,Reynolds:2016rvl,Chapman:2016hwi} with the cut-off distance from the AdS boundary in Fefferman-Graham coordinates identified as $\delta \sim 1/ \Lambda$. Eq.~\eqref{eq.LnNormsDivB} for $\mathcal{C}^{(1)}$ indicates  a leading divergence of $\text{Vol} \, \Lambda^{d} \left| \log  \frac{M}{\Lambda} \right|$ (with $M$ and $\Lambda$ independent), which resembles the result of the CA proposal.
In the holographic CA calculation, the leading logarithmic divergence is due to the codimension-$2$ joint action contributions associated with the intersection between the null and timelike hypersurfaces that bound the regulated Wheeler-DeWitt patch near the AdS boundary \cite{Lehner:2016vdi}.
These contributions depend on the parametrization of null normals (\cite{Lehner:2016vdi} suggested working in an affine parametrization) and their overall rescaling. 	
The latter gives rise to an extra freedom represented in \cite{Carmi:2016wjl} by a free parameter ${\tilde \alpha}$ inside the logarithm.
In our calculation the same type of ambiguity is related to the choice of the reference state scale $M$ and we can identify $M\sim {\tilde \alpha}/L_{\text{AdS}}$ where $L_{\text{AdS}}$ is the AdS scale.
When $M=\omega_{\Lambda}$, the leading divergence becomes proportional to  $\text{Vol} \, \Lambda^{d}$, which is in agreement with the CV results \cite{Carmi:2016wjl} (or with the CA results when including a counter term which renders the action
 reparametrization invariant, see \cite{Reynolds:2016rvl}).
It is interesting that despite considering QFTs without semiclassical gravity duals (having small central charge and no interactions), the $\mathcal{C}^{(1)}$ norm exhibits close similarity to the holographic calculations of leading UV divergences.

\vspace{5 pt}

\noindent \emph{7. Summary and Outlook.--}
We proposed a definition of state complexity in QFTs, independent from a notion of unitary complexity.
This measure is derived from the FS metric by restricting to  directions, in the space of states, generated by exponentiating
allowed generators $\cG$, on which our measure crucially depends.
We identified unitary paths that map simple Gaussian reference states $\ket{R(M)}$ with no spatial correlations to approximate ground states of free QFTs, generated within $\mathfrak{su}(1,1)$ subalgebras of momentum preserving quadratic generators and singled out the paths corresponding to minimal complexity according to our measure. Remarkably, for some instances, the evaluated complexity presents qualitative agreement with holographic results.

We could verify using our methods that cMERA circuits are optimal in the $\cC^{(1)}$ norm when interpreting the renormalization scale $u$ of cMERA as the circuit parameter $\sigma$.
In contrast, the $\cC^{(2)}$ norm allows  for lower FS complexity than that achieved by cMERA circuits by reorganizing the circuit in such a way that all the different momentum gates are active at every step along the circuit (see \cite{SupMat}A for details). The $C^{(1)}$ norm results show close resemblance to the holographic results which suggests  it is a better predictor of circuit complexity.

We worked in momentum space and restricted the generators to be quadratic. In position space our generators are bi-local which suggests an analogy to the 2-qubit operations of traditional quantum circuits. However our gates are spread in position space and it would be interesting to explore the implications of working with local gates.
Future directions include evaluating the complexity for fermionic systems
and studying the time evolution of  thermofield double states.
Finally it would be interesting to understand what universal data can be extracted from complexity, whether complexity in QFTs can serve as an order parameter, and if it plays a role in the context of RG-flows.

\begin{acknowledgments}
\section*{Acknowledgments}
We would like to express first our special gratitude to R.~C.~Myers for numerous illuminating discussions and suggestions, that helped shape the ideas and results of this paper and for sharing with us his preliminary results with R.~Jefferson  on defining complexity in QFTs using Nielsen's approach. S.~C. would like to express  personal gratitude for multiple discussions during Strings 2017 which motivated the development of the ideas of Sec.~5.
We are also particularly thankful to R.~Jefferson for many illuminating discussions and for providing comments on the supplementary material comparing the two works. We are also grateful to J.~Eisert for providing numerous comments on this work.
In addition, we would like to thank G.~Verdon-Akzam, J.~de~Boer, P.~Caputa, M.~Fleury, A.~Franco, K.~Hashimoto, Q.~Hu, R.~Janik, J.~Jottar, T.~Osborne, G.~Policastro,  K.~Rejzner, D.~Sarkar, V.~Scholz, M.~Spalinski, T.~Takayanagi, K.~Temme, J. Teschner, G.~Vidal, F.~Verstraete and P.~Witaszczyk for valuable comments and discussions.
Research at Perimeter Institute is supported by the Government of Canada through Industry Canada and by the Province of Ontario through the Ministry of Research \& Innovation. S.C. acknowledges additional support from an Israeli Women in Science Fellowship from the Israeli Council of Higher Education.
The research of M.P.H. is supported by the Alexander von Humboldt Foundation and the Federal Ministry for Education and Research through the Sofja Kovalevskaja Award.
M.P.H is also grateful to Perimeter Institute, ETH and the University of Amsterdam for stimulating hospitality during the completion of this project and to Kyoto University where this work was presented for the first time during the workshop \emph{Quantum Gravity, String Theory and Holography} in April 2017.
F.P. would also like to acknowledge the support of the Alexander von Humboldt Foundation.
\end{acknowledgments}

\bibliography{cMERA_complexity}{}

\begin{thebibliography}{43}%
\makeatletter
\providecommand \@ifxundefined [1]{%
 \@ifx{#1\undefined}
}%
\providecommand \@ifnum [1]{%
 \ifnum #1\expandafter \@firstoftwo
 \else \expandafter \@secondoftwo
 \fi
}%
\providecommand \@ifx [1]{%
 \ifx #1\expandafter \@firstoftwo
 \else \expandafter \@secondoftwo
 \fi
}%
\providecommand \natexlab [1]{#1}%
\providecommand \enquote  [1]{``#1''}%
\providecommand \bibnamefont  [1]{#1}%
\providecommand \bibfnamefont [1]{#1}%
\providecommand \citenamefont [1]{#1}%
\providecommand \href@noop [0]{\@secondoftwo}%
\providecommand \href [0]{\begingroup \@sanitize@url \@href}%
\providecommand \@href[1]{\@@startlink{#1}\@@href}%
\providecommand \@@href[1]{\endgroup#1\@@endlink}%
\providecommand \@sanitize@url [0]{\catcode `\\12\catcode `\$12\catcode
  `\&12\catcode `\#12\catcode `\^12\catcode `\_12\catcode `\%12\relax}%
\providecommand \@@startlink[1]{}%
\providecommand \@@endlink[0]{}%
\providecommand \url  [0]{\begingroup\@sanitize@url \@url }%
\providecommand \@url [1]{\endgroup\@href {#1}{\urlprefix }}%
\providecommand \urlprefix  [0]{URL }%
\providecommand \Eprint [0]{\href }%
\providecommand \doibase [0]{http://dx.doi.org/}%
\providecommand \selectlanguage [0]{\@gobble}%
\providecommand \bibinfo  [0]{\@secondoftwo}%
\providecommand \bibfield  [0]{\@secondoftwo}%
\providecommand \translation [1]{[#1]}%
\providecommand \BibitemOpen [0]{}%
\providecommand \bibitemStop [0]{}%
\providecommand \bibitemNoStop [0]{.\EOS\space}%
\providecommand \EOS [0]{\spacefactor3000\relax}%
\providecommand \BibitemShut  [1]{\csname bibitem#1\endcsname}%
\let\auto@bib@innerbib\@empty
\bibitem [{\citenamefont {Maldacena}(1999)}]{Maldacena:1997re}%
  \BibitemOpen
  \bibfield  {author} {\bibinfo {author} {\bibfnamefont {J.~M.}\ \bibnamefont
  {Maldacena}},\ }\href {\doibase 10.1023/A:1026654312961} {\bibfield
  {journal} {\bibinfo  {journal} {Int. J. Theor. Phys.}\ }\textbf {\bibinfo
  {volume} {38}},\ \bibinfo {pages} {1113} (\bibinfo {year} {1999})},\ \bibinfo
  {note} {[Adv. Theor. Math. Phys.2,231(1998)]},\ \Eprint
  {http://arxiv.org/abs/hep-th/9711200} {arXiv:hep-th/9711200 [hep-th]}
  \BibitemShut {NoStop}%
\bibitem [{\citenamefont {Van~Raamsdonk}(2010)}]{VanRaamsdonk:2010pw}%
  \BibitemOpen
  \bibfield  {author} {\bibinfo {author} {\bibfnamefont {M.}~\bibnamefont
  {Van~Raamsdonk}},\ }\href {\doibase 10.1007/s10714-010-1034-0,
  10.1142/S0218271810018529} {\bibfield  {journal} {\bibinfo  {journal} {Gen.
  Rel. Grav.}\ }\textbf {\bibinfo {volume} {42}},\ \bibinfo {pages} {2323}
  (\bibinfo {year} {2010})},\ \bibinfo {note} {[Int. J. Mod.
  Phys.D19,2429(2010)]},\ \Eprint {http://arxiv.org/abs/1005.3035}
  {arXiv:1005.3035 [hep-th]} \BibitemShut {NoStop}%
\bibitem [{\citenamefont {Ryu}\ and\ \citenamefont
  {Takayanagi}(2006)}]{Ryu:2006bv}%
  \BibitemOpen
  \bibfield  {author} {\bibinfo {author} {\bibfnamefont {S.}~\bibnamefont
  {Ryu}}\ and\ \bibinfo {author} {\bibfnamefont {T.}~\bibnamefont
  {Takayanagi}},\ }\href {\doibase 10.1103/PhysRevLett.96.181602} {\bibfield
  {journal} {\bibinfo  {journal} {Phys. Rev. Lett.}\ }\textbf {\bibinfo
  {volume} {96}},\ \bibinfo {pages} {181602} (\bibinfo {year} {2006})},\
  \Eprint {http://arxiv.org/abs/hep-th/0603001} {arXiv:hep-th/0603001 [hep-th]}
  \BibitemShut {NoStop}%
\bibitem [{\citenamefont {Rangamani}\ and\ \citenamefont
  {Takayanagi}(2016)}]{Rangamani:2016dms}%
  \BibitemOpen
  \bibfield  {author} {\bibinfo {author} {\bibfnamefont {M.}~\bibnamefont
  {Rangamani}}\ and\ \bibinfo {author} {\bibfnamefont {T.}~\bibnamefont
  {Takayanagi}},\ }\href@noop {} {\  (\bibinfo {year} {2016})},\ \Eprint
  {http://arxiv.org/abs/1609.01287} {arXiv:1609.01287 [hep-th]} \BibitemShut
  {NoStop}%
\bibitem [{\citenamefont {Balasubramanian}\ \emph {et~al.}(2015)\citenamefont
  {Balasubramanian}, \citenamefont {Chowdhury}, \citenamefont {Czech},\ and\
  \citenamefont {de~Boer}}]{Balasubramanian:2014sra}%
  \BibitemOpen
  \bibfield  {author} {\bibinfo {author} {\bibfnamefont {V.}~\bibnamefont
  {Balasubramanian}}, \bibinfo {author} {\bibfnamefont {B.~D.}\ \bibnamefont
  {Chowdhury}}, \bibinfo {author} {\bibfnamefont {B.}~\bibnamefont {Czech}}, \
  and\ \bibinfo {author} {\bibfnamefont {J.}~\bibnamefont {de~Boer}},\ }\href
  {\doibase 10.1007/JHEP01(2015)048} {\bibfield  {journal} {\bibinfo  {journal}
  {JHEP}\ }\textbf {\bibinfo {volume} {01}},\ \bibinfo {pages} {048} (\bibinfo
  {year} {2015})},\ \Eprint {http://arxiv.org/abs/1406.5859} {arXiv:1406.5859
  [hep-th]} \BibitemShut {NoStop}%
\bibitem [{\citenamefont {Susskind}(2016{\natexlab{a}})}]{Susskind:2014moa}%
  \BibitemOpen
  \bibfield  {author} {\bibinfo {author} {\bibfnamefont {L.}~\bibnamefont
  {Susskind}},\ }\href {\doibase 10.1002/prop.201500095} {\bibfield  {journal}
  {\bibinfo  {journal} {Fortsch. Phys.}\ }\textbf {\bibinfo {volume} {64}},\
  \bibinfo {pages} {49} (\bibinfo {year} {2016}{\natexlab{a}})},\ \Eprint
  {http://arxiv.org/abs/1411.0690} {arXiv:1411.0690 [hep-th]} \BibitemShut
  {NoStop}%
\bibitem [{\citenamefont {Freivogel}\ \emph {et~al.}(2015)\citenamefont
  {Freivogel}, \citenamefont {Jefferson}, \citenamefont {Kabir}, \citenamefont
  {Mosk},\ and\ \citenamefont {Yang}}]{Freivogel:2014lja}%
  \BibitemOpen
  \bibfield  {author} {\bibinfo {author} {\bibfnamefont {B.}~\bibnamefont
  {Freivogel}}, \bibinfo {author} {\bibfnamefont {R.~A.}\ \bibnamefont
  {Jefferson}}, \bibinfo {author} {\bibfnamefont {L.}~\bibnamefont {Kabir}},
  \bibinfo {author} {\bibfnamefont {B.}~\bibnamefont {Mosk}}, \ and\ \bibinfo
  {author} {\bibfnamefont {I.-S.}\ \bibnamefont {Yang}},\ }\href {\doibase
  10.1103/PhysRevD.91.086013} {\bibfield  {journal} {\bibinfo  {journal} {Phys.
  Rev.}\ }\textbf {\bibinfo {volume} {D91}},\ \bibinfo {pages} {086013}
  (\bibinfo {year} {2015})},\ \Eprint {http://arxiv.org/abs/1412.5175}
  {arXiv:1412.5175 [hep-th]} \BibitemShut {NoStop}%
\bibitem [{\citenamefont {Susskind}(2016{\natexlab{b}})}]{Susskind:2014rva}%
  \BibitemOpen
  \bibfield  {author} {\bibinfo {author} {\bibfnamefont {L.}~\bibnamefont
  {Susskind}},\ }\href {\doibase 10.1002/prop.201500092} {\bibfield  {journal}
  {\bibinfo  {journal} {Fortsch. Phys.}\ }\textbf {\bibinfo {volume} {64}},\
  \bibinfo {pages} {24} (\bibinfo {year} {2016}{\natexlab{b}})},\ \Eprint
  {http://arxiv.org/abs/1403.5695} {arXiv:1403.5695 [hep-th]} \BibitemShut
  {NoStop}%
\bibitem [{\citenamefont {Stanford}\ and\ \citenamefont
  {Susskind}(2014)}]{Stanford:2014jda}%
  \BibitemOpen
  \bibfield  {author} {\bibinfo {author} {\bibfnamefont {D.}~\bibnamefont
  {Stanford}}\ and\ \bibinfo {author} {\bibfnamefont {L.}~\bibnamefont
  {Susskind}},\ }\href {\doibase 10.1103/PhysRevD.90.126007} {\bibfield
  {journal} {\bibinfo  {journal} {Phys. Rev.}\ }\textbf {\bibinfo {volume}
  {D90}},\ \bibinfo {pages} {126007} (\bibinfo {year} {2014})},\ \Eprint
  {http://arxiv.org/abs/1406.2678} {arXiv:1406.2678 [hep-th]} \BibitemShut
  {NoStop}%
\bibitem [{\citenamefont {Alishahiha}(2015)}]{Alishahiha:2015rta}%
  \BibitemOpen
  \bibfield  {author} {\bibinfo {author} {\bibfnamefont {M.}~\bibnamefont
  {Alishahiha}},\ }\href {\doibase 10.1103/PhysRevD.92.126009} {\bibfield
  {journal} {\bibinfo  {journal} {Phys. Rev.}\ }\textbf {\bibinfo {volume}
  {D92}},\ \bibinfo {pages} {126009} (\bibinfo {year} {2015})},\ \Eprint
  {http://arxiv.org/abs/1509.06614} {arXiv:1509.06614 [hep-th]} \BibitemShut
  {NoStop}%
\bibitem [{\citenamefont {Brown}\ \emph
  {et~al.}(2016{\natexlab{a}})\citenamefont {Brown}, \citenamefont {Roberts},
  \citenamefont {Susskind}, \citenamefont {Swingle},\ and\ \citenamefont
  {Zhao}}]{Brown:2015bva}%
  \BibitemOpen
  \bibfield  {author} {\bibinfo {author} {\bibfnamefont {A.~R.}\ \bibnamefont
  {Brown}}, \bibinfo {author} {\bibfnamefont {D.~A.}\ \bibnamefont {Roberts}},
  \bibinfo {author} {\bibfnamefont {L.}~\bibnamefont {Susskind}}, \bibinfo
  {author} {\bibfnamefont {B.}~\bibnamefont {Swingle}}, \ and\ \bibinfo
  {author} {\bibfnamefont {Y.}~\bibnamefont {Zhao}},\ }\href {\doibase
  10.1103/PhysRevLett.116.191301} {\bibfield  {journal} {\bibinfo  {journal}
  {Phys. Rev. Lett.}\ }\textbf {\bibinfo {volume} {116}},\ \bibinfo {pages}
  {191301} (\bibinfo {year} {2016}{\natexlab{a}})},\ \Eprint
  {http://arxiv.org/abs/1509.07876} {arXiv:1509.07876 [hep-th]} \BibitemShut
  {NoStop}%
\bibitem [{\citenamefont {Brown}\ \emph
  {et~al.}(2016{\natexlab{b}})\citenamefont {Brown}, \citenamefont {Roberts},
  \citenamefont {Susskind}, \citenamefont {Swingle},\ and\ \citenamefont
  {Zhao}}]{Brown:2015lvg}%
  \BibitemOpen
  \bibfield  {author} {\bibinfo {author} {\bibfnamefont {A.~R.}\ \bibnamefont
  {Brown}}, \bibinfo {author} {\bibfnamefont {D.~A.}\ \bibnamefont {Roberts}},
  \bibinfo {author} {\bibfnamefont {L.}~\bibnamefont {Susskind}}, \bibinfo
  {author} {\bibfnamefont {B.}~\bibnamefont {Swingle}}, \ and\ \bibinfo
  {author} {\bibfnamefont {Y.}~\bibnamefont {Zhao}},\ }\href {\doibase
  10.1103/PhysRevD.93.086006} {\bibfield  {journal} {\bibinfo  {journal} {Phys.
  Rev.}\ }\textbf {\bibinfo {volume} {D93}},\ \bibinfo {pages} {086006}
  (\bibinfo {year} {2016}{\natexlab{b}})},\ \Eprint
  {http://arxiv.org/abs/1512.04993} {arXiv:1512.04993 [hep-th]} \BibitemShut
  {NoStop}%
\bibitem [{\citenamefont {Carmi}\ \emph {et~al.}(2017)\citenamefont {Carmi},
  \citenamefont {Myers},\ and\ \citenamefont {Rath}}]{Carmi:2016wjl}%
  \BibitemOpen
  \bibfield  {author} {\bibinfo {author} {\bibfnamefont {D.}~\bibnamefont
  {Carmi}}, \bibinfo {author} {\bibfnamefont {R.~C.}\ \bibnamefont {Myers}}, \
  and\ \bibinfo {author} {\bibfnamefont {P.}~\bibnamefont {Rath}},\ }\href
  {\doibase 10.1007/JHEP03(2017)118} {\bibfield  {journal} {\bibinfo  {journal}
  {JHEP}\ }\textbf {\bibinfo {volume} {03}},\ \bibinfo {pages} {118} (\bibinfo
  {year} {2017})},\ \Eprint {http://arxiv.org/abs/1612.00433} {arXiv:1612.00433
  [hep-th]} \BibitemShut {NoStop}%
\bibitem [{\citenamefont {Chapman}\ \emph {et~al.}(2017)\citenamefont
  {Chapman}, \citenamefont {Marrochio},\ and\ \citenamefont
  {Myers}}]{Chapman:2016hwi}%
  \BibitemOpen
  \bibfield  {author} {\bibinfo {author} {\bibfnamefont {S.}~\bibnamefont
  {Chapman}}, \bibinfo {author} {\bibfnamefont {H.}~\bibnamefont {Marrochio}},
  \ and\ \bibinfo {author} {\bibfnamefont {R.~C.}\ \bibnamefont {Myers}},\
  }\href {\doibase 10.1007/JHEP01(2017)062} {\bibfield  {journal} {\bibinfo
  {journal} {JHEP}\ }\textbf {\bibinfo {volume} {01}},\ \bibinfo {pages} {062}
  (\bibinfo {year} {2017})},\ \Eprint {http://arxiv.org/abs/1610.08063}
  {arXiv:1610.08063 [hep-th]} \BibitemShut {NoStop}%
\bibitem [{\citenamefont {Toffoli}(1998)}]{Toffoli1998}%
  \BibitemOpen
  \bibfield  {author} {\bibinfo {author} {\bibfnamefont {T.}~\bibnamefont
  {Toffoli}},\ }\href {\doibase 10.1016/S0167-2789(98)00040-2} {\bibfield
  {journal} {\bibinfo  {journal} {Physica D: Nonlinear Phenomena}\ }\textbf
  {\bibinfo {volume} {120}},\ \bibinfo {pages} {1} (\bibinfo {year}
  {1998})}\BibitemShut {NoStop}%
\bibitem [{\citenamefont {Toffoli}(1999)}]{Toffoli1999}%
  \BibitemOpen
  \bibfield  {author} {\bibinfo {author} {\bibfnamefont {T.}~\bibnamefont
  {Toffoli}},\ }in\ \href {http://dl.acm.org/citation.cfm?id=304763} {\emph
  {\bibinfo {booktitle} {{Feynman and computation : exploring the limits of
  computers}}}},\ \bibinfo {editor} {edited by\ \bibinfo {editor}
  {\bibfnamefont {A.~J.~G.}\ \bibnamefont {Hey}}\ and\ \bibinfo {editor}
  {\bibfnamefont {R.~P.}\ \bibnamefont {Feynman}}}\ (\bibinfo  {publisher}
  {Perseus Books},\ \bibinfo {year} {1999})\ pp.\ \bibinfo {pages}
  {349--392}\BibitemShut {NoStop}%
\bibitem [{\citenamefont {Nielsen}\ \emph {et~al.}(2006)\citenamefont
  {Nielsen}, \citenamefont {Dowling}, \citenamefont {Gu},\ and\ \citenamefont
  {Doherty}}]{Nielsen2006}%
  \BibitemOpen
  \bibfield  {author} {\bibinfo {author} {\bibfnamefont {M.~A.}\ \bibnamefont
  {Nielsen}}, \bibinfo {author} {\bibfnamefont {M.~R.}\ \bibnamefont
  {Dowling}}, \bibinfo {author} {\bibfnamefont {M.}~\bibnamefont {Gu}}, \ and\
  \bibinfo {author} {\bibfnamefont {A.~C.}\ \bibnamefont {Doherty}},\ }\href
  {http://science.sciencemag.org/content/311/5764/1133.full?ijkey=SmuV2ij{\&}keytype=ref{\&}siteid=sci}
  {\bibfield  {journal} {\bibinfo  {journal} {Science}\ }\textbf {\bibinfo
  {volume} {311}} (\bibinfo {year} {2006})}\BibitemShut {NoStop}%
\bibitem [{\citenamefont {{Watrous}}(2008)}]{WatrousComp}%
  \BibitemOpen
  \bibfield  {author} {\bibinfo {author} {\bibfnamefont {J.}~\bibnamefont
  {{Watrous}}},\ }\href@noop {} {\bibfield  {journal} {\bibinfo  {journal}
  {ArXiv e-prints}\ } (\bibinfo {year} {2008})},\ \Eprint
  {http://arxiv.org/abs/0804.3401} {arXiv:0804.3401 [quant-ph]} \BibitemShut
  {NoStop}%
\bibitem [{\citenamefont {Aaronson}(2016)}]{Aaronson:2016vto}%
  \BibitemOpen
  \bibfield  {author} {\bibinfo {author} {\bibfnamefont {S.}~\bibnamefont
  {Aaronson}}\ }(\bibinfo {year} {2016})\ \Eprint
  {http://arxiv.org/abs/1607.05256} {arXiv:1607.05256 [quant-ph]} \BibitemShut
  {NoStop}%
\bibitem [{\citenamefont {Swingle}(2012)}]{Swingle2012}%
  \BibitemOpen
  \bibfield  {author} {\bibinfo {author} {\bibfnamefont {B.}~\bibnamefont
  {Swingle}},\ }\href {\doibase 10.1103/PhysRevD.86.065007} {\bibfield
  {journal} {\bibinfo  {journal} {Physical Review D}\ }\textbf {\bibinfo
  {volume} {86}},\ \bibinfo {pages} {065007} (\bibinfo {year}
  {2012})}\BibitemShut {NoStop}%
\bibitem [{\citenamefont {Vidal}(2008)}]{MERA}%
  \BibitemOpen
  \bibfield  {author} {\bibinfo {author} {\bibfnamefont {G.}~\bibnamefont
  {Vidal}},\ }\href {\doibase 10.1103/PhysRevLett.101.110501} {\bibfield
  {journal} {\bibinfo  {journal} {Phys. Rev. Lett.}\ }\textbf {\bibinfo
  {volume} {101}},\ \bibinfo {pages} {110501} (\bibinfo {year}
  {2008})}\BibitemShut {NoStop}%
\bibitem [{\citenamefont {Jefferson}\ and\ \citenamefont
  {Myers}(2017)}]{Jefferson:2017sdb}%
  \BibitemOpen
  \bibfield  {author} {\bibinfo {author} {\bibfnamefont {R.~A.}\ \bibnamefont
  {Jefferson}}\ and\ \bibinfo {author} {\bibfnamefont {R.~C.}\ \bibnamefont
  {Myers}},\ }\href@noop {} {\  (\bibinfo {year} {2017})},\ \Eprint
  {http://arxiv.org/abs/1707.08570} {arXiv:1707.08570 [hep-th]} \BibitemShut
  {NoStop}%
\bibitem [{Sup()}]{SupMat}%
  \BibitemOpen
  \href@noop {} {}\bibinfo {note} {Supplemental material, composed of six
  sections. Section A evaluates the length of the path traced by the cMERA
  circuit according to the FS metric. Section B explores the properties of
  ground state complexities $\mathcal{C}^{(n)}$ and evaluates the complexity of
  massive field theory ground states with respect to the CFT vacuum state.
  Section C provides identifies hyperbolic planes in the FS manifold induced by
  the $\mathfrak{su}(1,1)$ algebras. Section D considers a lower bounded
  constant generator which induces constant period oscillations between the
  reference and target states and evaluate the corresponding FS length. Section
  E presents a simplified derivation of the hyperbolic plane metric for a
  single pair of modes. Section F explores similarities and differences between
  this work and the work by Jefferson and Myers.}\BibitemShut {Stop}%
\bibitem [{\citenamefont {Haegeman}\ \emph {et~al.}(2013)\citenamefont
  {Haegeman}, \citenamefont {Osborne}, \citenamefont {Verschelde},\ and\
  \citenamefont {Verstraete}}]{Haegeman:2011uy}%
  \BibitemOpen
  \bibfield  {author} {\bibinfo {author} {\bibfnamefont {J.}~\bibnamefont
  {Haegeman}}, \bibinfo {author} {\bibfnamefont {T.~J.}\ \bibnamefont
  {Osborne}}, \bibinfo {author} {\bibfnamefont {H.}~\bibnamefont {Verschelde}},
  \ and\ \bibinfo {author} {\bibfnamefont {F.}~\bibnamefont {Verstraete}},\
  }\href {\doibase 10.1103/PhysRevLett.110.100402} {\bibfield  {journal}
  {\bibinfo  {journal} {Phys. Rev. Lett.}\ }\textbf {\bibinfo {volume} {110}},\
  \bibinfo {pages} {100402} (\bibinfo {year} {2013})},\ \Eprint
  {http://arxiv.org/abs/1102.5524} {arXiv:1102.5524 [hep-th]} \BibitemShut
  {NoStop}%
\bibitem [{\citenamefont {Nozaki}\ \emph {et~al.}(2012)\citenamefont {Nozaki},
  \citenamefont {Ryu},\ and\ \citenamefont {Takayanagi}}]{Nozaki:2012zj}%
  \BibitemOpen
  \bibfield  {author} {\bibinfo {author} {\bibfnamefont {M.}~\bibnamefont
  {Nozaki}}, \bibinfo {author} {\bibfnamefont {S.}~\bibnamefont {Ryu}}, \ and\
  \bibinfo {author} {\bibfnamefont {T.}~\bibnamefont {Takayanagi}},\ }\href
  {\doibase 10.1007/JHEP10(2012)193} {\bibfield  {journal} {\bibinfo  {journal}
  {JHEP}\ }\textbf {\bibinfo {volume} {10}},\ \bibinfo {pages} {193} (\bibinfo
  {year} {2012})},\ \Eprint {http://arxiv.org/abs/1208.3469} {arXiv:1208.3469
  [hep-th]} \BibitemShut {NoStop}%
\bibitem [{\citenamefont {Mollabashi}\ \emph {et~al.}(2014)\citenamefont
  {Mollabashi}, \citenamefont {Nozaki}, \citenamefont {Ryu},\ and\
  \citenamefont {Takayanagi}}]{Mollabashi:2013lya}%
  \BibitemOpen
  \bibfield  {author} {\bibinfo {author} {\bibfnamefont {A.}~\bibnamefont
  {Mollabashi}}, \bibinfo {author} {\bibfnamefont {M.}~\bibnamefont {Nozaki}},
  \bibinfo {author} {\bibfnamefont {S.}~\bibnamefont {Ryu}}, \ and\ \bibinfo
  {author} {\bibfnamefont {T.}~\bibnamefont {Takayanagi}},\ }\href {\doibase
  10.1007/JHEP03(2014)098} {\bibfield  {journal} {\bibinfo  {journal} {JHEP}\
  }\textbf {\bibinfo {volume} {03}},\ \bibinfo {pages} {098} (\bibinfo {year}
  {2014})},\ \Eprint {http://arxiv.org/abs/1311.6095} {arXiv:1311.6095
  [hep-th]} \BibitemShut {NoStop}%
\bibitem [{\citenamefont {Miyaji}\ and\ \citenamefont
  {Takayanagi}(2015)}]{Miyaji2015}%
  \BibitemOpen
  \bibfield  {author} {\bibinfo {author} {\bibfnamefont {M.}~\bibnamefont
  {Miyaji}}\ and\ \bibinfo {author} {\bibfnamefont {T.}~\bibnamefont
  {Takayanagi}},\ }\href {\doibase 10.1093/ptep/ptv089} {\  (\bibinfo {year}
  {2015}),\ 10.1093/ptep/ptv089},\ \Eprint {http://arxiv.org/abs/1503.03542}
  {arXiv:1503.03542} \BibitemShut {NoStop}%
\bibitem [{\citenamefont {Molina-Vilaplana}(2015)}]{Molina-Vilaplana2015}%
  \BibitemOpen
  \bibfield  {author} {\bibinfo {author} {\bibfnamefont {J.}~\bibnamefont
  {Molina-Vilaplana}},\ }\href {\doibase 10.1007/JHEP09(2015)002} {\bibfield
  {journal} {\bibinfo  {journal} {Journal of High Energy Physics}\ }\textbf
  {\bibinfo {volume} {2015}},\ \bibinfo {pages} {2} (\bibinfo {year}
  {2015})}\BibitemShut {NoStop}%
\bibitem [{\citenamefont {Caputa}\ \emph
  {et~al.}(2017{\natexlab{a}})\citenamefont {Caputa}, \citenamefont {Kundu},
  \citenamefont {Miyaji}, \citenamefont {Takayanagi},\ and\ \citenamefont
  {Watanabe}}]{Caputa:2017urj}%
  \BibitemOpen
  \bibfield  {author} {\bibinfo {author} {\bibfnamefont {P.}~\bibnamefont
  {Caputa}}, \bibinfo {author} {\bibfnamefont {N.}~\bibnamefont {Kundu}},
  \bibinfo {author} {\bibfnamefont {M.}~\bibnamefont {Miyaji}}, \bibinfo
  {author} {\bibfnamefont {T.}~\bibnamefont {Takayanagi}}, \ and\ \bibinfo
  {author} {\bibfnamefont {K.}~\bibnamefont {Watanabe}},\ }\href@noop {} {\
  (\bibinfo {year} {2017}{\natexlab{a}})},\ \Eprint
  {http://arxiv.org/abs/1703.00456} {arXiv:1703.00456 [hep-th]} \BibitemShut
  {NoStop}%
\bibitem [{\citenamefont {Caputa}\ \emph
  {et~al.}(2017{\natexlab{b}})\citenamefont {Caputa}, \citenamefont {Kundu},
  \citenamefont {Miyaji}, \citenamefont {Takayanagi},\ and\ \citenamefont
  {Watanabe}}]{Caputa:2017yrh}%
  \BibitemOpen
  \bibfield  {author} {\bibinfo {author} {\bibfnamefont {P.}~\bibnamefont
  {Caputa}}, \bibinfo {author} {\bibfnamefont {N.}~\bibnamefont {Kundu}},
  \bibinfo {author} {\bibfnamefont {M.}~\bibnamefont {Miyaji}}, \bibinfo
  {author} {\bibfnamefont {T.}~\bibnamefont {Takayanagi}}, \ and\ \bibinfo
  {author} {\bibfnamefont {K.}~\bibnamefont {Watanabe}},\ }\href@noop {} {\
  (\bibinfo {year} {2017}{\natexlab{b}})},\ \Eprint
  {http://arxiv.org/abs/1706.07056} {arXiv:1706.07056 [hep-th]} \BibitemShut
  {NoStop}%
\bibitem [{\citenamefont {Czech}(2017)}]{Czech2017}%
  \BibitemOpen
  \bibfield  {author} {\bibinfo {author} {\bibfnamefont {B.}~\bibnamefont
  {Czech}},\ }\href {http://arxiv.org/abs/1706.00965} {\  (\bibinfo {year}
  {2017})},\ \Eprint {http://arxiv.org/abs/1706.00965} {arXiv:1706.00965}
  \BibitemShut {NoStop}%
\bibitem [{\citenamefont {Brown}\ and\ \citenamefont
  {Susskind}(2017)}]{Brown:2017jil}%
  \BibitemOpen
  \bibfield  {author} {\bibinfo {author} {\bibfnamefont {A.~R.}\ \bibnamefont
  {Brown}}\ and\ \bibinfo {author} {\bibfnamefont {L.}~\bibnamefont
  {Susskind}},\ }\href@noop {} {\  (\bibinfo {year} {2017})},\ \Eprint
  {http://arxiv.org/abs/1701.01107} {arXiv:1701.01107 [hep-th]} \BibitemShut
  {NoStop}%
\bibitem [{\citenamefont {Hashimoto}\ \emph {et~al.}(2017)\citenamefont
  {Hashimoto}, \citenamefont {Iizuka},\ and\ \citenamefont
  {Sugishita}}]{Hashimoto:2017fga}%
  \BibitemOpen
  \bibfield  {author} {\bibinfo {author} {\bibfnamefont {K.}~\bibnamefont
  {Hashimoto}}, \bibinfo {author} {\bibfnamefont {N.}~\bibnamefont {Iizuka}}, \
  and\ \bibinfo {author} {\bibfnamefont {S.}~\bibnamefont {Sugishita}},\
  }\href@noop {} {\  (\bibinfo {year} {2017})},\ \Eprint
  {http://arxiv.org/abs/1707.03840} {arXiv:1707.03840 [hep-th]} \BibitemShut
  {NoStop}%
\bibitem [{\citenamefont {Roberts}\ and\ \citenamefont
  {Yoshida}(2017)}]{Roberts2017}%
  \BibitemOpen
  \bibfield  {author} {\bibinfo {author} {\bibfnamefont {D.~A.}\ \bibnamefont
  {Roberts}}\ and\ \bibinfo {author} {\bibfnamefont {B.}~\bibnamefont
  {Yoshida}},\ }\href {\doibase 10.1007/JHEP04(2017)121} {\bibfield  {journal}
  {\bibinfo  {journal} {Journal of High Energy Physics}\ }\textbf {\bibinfo
  {volume} {2017}},\ \bibinfo {pages} {121} (\bibinfo {year}
  {2017})}\BibitemShut {NoStop}%
\bibitem [{\citenamefont {Chemissany}\ and\ \citenamefont
  {Osborne}(2016)}]{Chemissany:2016qqq}%
  \BibitemOpen
  \bibfield  {author} {\bibinfo {author} {\bibfnamefont {W.}~\bibnamefont
  {Chemissany}}\ and\ \bibinfo {author} {\bibfnamefont {T.~J.}\ \bibnamefont
  {Osborne}},\ }\href {\doibase 10.1007/JHEP12(2016)055} {\bibfield  {journal}
  {\bibinfo  {journal} {JHEP}\ }\textbf {\bibinfo {volume} {12}},\ \bibinfo
  {pages} {055} (\bibinfo {year} {2016})},\ \Eprint
  {http://arxiv.org/abs/1605.07768} {arXiv:1605.07768 [hep-th]} \BibitemShut
  {NoStop}%
\bibitem [{\citenamefont {Bengtsson}\ and\ \citenamefont
  {Zyczkowski}(2007)}]{bengtsson2007geometry}%
  \BibitemOpen
  \bibfield  {author} {\bibinfo {author} {\bibfnamefont {I.}~\bibnamefont
  {Bengtsson}}\ and\ \bibinfo {author} {\bibfnamefont {K.}~\bibnamefont
  {Zyczkowski}},\ }\href {https://books.google.de/books?id=aA4vXMbuOTUC} {\emph
  {\bibinfo {title} {Geometry of Quantum States: An Introduction to Quantum
  Entanglement}}}\ (\bibinfo  {publisher} {Cambridge University Press},\
  \bibinfo {year} {2007})\BibitemShut {NoStop}%
\bibitem [{\citenamefont {Hu}\ and\ \citenamefont {Vidal}(2017)}]{Hu:2017rsp}%
  \BibitemOpen
  \bibfield  {author} {\bibinfo {author} {\bibfnamefont {Q.}~\bibnamefont
  {Hu}}\ and\ \bibinfo {author} {\bibfnamefont {G.}~\bibnamefont {Vidal}},\
  }\href@noop {} {\  (\bibinfo {year} {2017})},\ \Eprint
  {http://arxiv.org/abs/1703.04798} {arXiv:1703.04798 [quant-ph]} \BibitemShut
  {NoStop}%
\bibitem [{\citenamefont {Perelomov}(1972)}]{Perelomov:1971bd}%
  \BibitemOpen
  \bibfield  {author} {\bibinfo {author} {\bibfnamefont {A.~M.}\ \bibnamefont
  {Perelomov}},\ }\href {\doibase 10.1007/BF01645091} {\bibfield  {journal}
  {\bibinfo  {journal} {Commun. Math. Phys.}\ }\textbf {\bibinfo {volume}
  {26}},\ \bibinfo {pages} {222} (\bibinfo {year} {1972})}\BibitemShut
  {NoStop}%
\bibitem [{\citenamefont {Provost}\ and\ \citenamefont
  {Vallee}(1980)}]{Provost:1980nc}%
  \BibitemOpen
  \bibfield  {author} {\bibinfo {author} {\bibfnamefont {J.~P.}\ \bibnamefont
  {Provost}}\ and\ \bibinfo {author} {\bibfnamefont {G.}~\bibnamefont
  {Vallee}},\ }\href {\doibase 10.1007/BF02193559} {\bibfield  {journal}
  {\bibinfo  {journal} {Commun. Math. Phys.}\ }\textbf {\bibinfo {volume}
  {76}},\ \bibinfo {pages} {289} (\bibinfo {year} {1980})}\BibitemShut
  {NoStop}%
\bibitem [{\citenamefont {Perelomov}(1986)}]{Perelomov:1986tf}%
  \BibitemOpen
  \bibfield  {author} {\bibinfo {author} {\bibfnamefont {A.~M.}\ \bibnamefont
  {Perelomov}},\ }\href@noop {} {\emph {\bibinfo {title} {{Generalized coherent
  states and their applications}}}}\ (\bibinfo {year} {1986})\BibitemShut
  {NoStop}%
\bibitem [{\citenamefont {Reynolds}\ and\ \citenamefont
  {Ross}(2017)}]{Reynolds:2016rvl}%
  \BibitemOpen
  \bibfield  {author} {\bibinfo {author} {\bibfnamefont {A.}~\bibnamefont
  {Reynolds}}\ and\ \bibinfo {author} {\bibfnamefont {S.~F.}\ \bibnamefont
  {Ross}},\ }\href {\doibase 10.1088/1361-6382/aa6925} {\bibfield  {journal}
  {\bibinfo  {journal} {Class. Quant. Grav.}\ }\textbf {\bibinfo {volume}
  {34}},\ \bibinfo {pages} {105004} (\bibinfo {year} {2017})},\ \Eprint
  {http://arxiv.org/abs/1612.05439} {arXiv:1612.05439 [hep-th]} \BibitemShut
  {NoStop}%
\bibitem [{\citenamefont {Lehner}\ \emph {et~al.}(2016)\citenamefont {Lehner},
  \citenamefont {Myers}, \citenamefont {Poisson},\ and\ \citenamefont
  {Sorkin}}]{Lehner:2016vdi}%
  \BibitemOpen
  \bibfield  {author} {\bibinfo {author} {\bibfnamefont {L.}~\bibnamefont
  {Lehner}}, \bibinfo {author} {\bibfnamefont {R.~C.}\ \bibnamefont {Myers}},
  \bibinfo {author} {\bibfnamefont {E.}~\bibnamefont {Poisson}}, \ and\
  \bibinfo {author} {\bibfnamefont {R.~D.}\ \bibnamefont {Sorkin}},\ }\href
  {\doibase 10.1103/PhysRevD.94.084046} {\bibfield  {journal} {\bibinfo
  {journal} {Phys. Rev.}\ }\textbf {\bibinfo {volume} {D94}},\ \bibinfo {pages}
  {084046} (\bibinfo {year} {2016})},\ \Eprint
  {http://arxiv.org/abs/1609.00207} {arXiv:1609.00207 [hep-th]} \BibitemShut
  {NoStop}%
\bibitem [{\citenamefont {Klimov}\ and\ \citenamefont
  {Chumakov}(2009)}]{Klimov2009}%
  \BibitemOpen
  \bibfield  {author} {\bibinfo {author} {\bibfnamefont {A.~B.}\ \bibnamefont
  {Klimov}}\ and\ \bibinfo {author} {\bibfnamefont {S.~M.}\ \bibnamefont
  {Chumakov}},\ }\href@noop {} {\emph {\bibinfo {title} {{A group-theoretical
  approach to quantum optics : models of atom-field interactions}}}}\ (\bibinfo
   {publisher} {Wiley-VCH},\ \bibinfo {year} {2009})\ p.\ \bibinfo {pages}
  {322}\BibitemShut {NoStop}%
\end{thebibliography}%

\appendix{}

\titlepage
\pagebreak

\setcounter{page}{1}

\begin{center}
{\LARGE Supplemental Material}
\end{center}
\section{A. cMERA circuit length according to the Fubini-Study metric}
Here, we evaluate the cMERA circuit length according to the proposed Fubini-Study metric.
We demonstrate that in the $\cC^{(2)}$ norm, the cMERA circuit is longer (i.e., more complex) than the minimal circuit described in the main text whereas in the $\cC^{(1)}$ norm its circuit length coincides with the corresponding minimal complexity.
We review below the needed ingredients of cMERA  and refer the reader to Ref.~\cite{Haegeman:2011uy,Nozaki:2012zj,Mollabashi:2013lya} for extra details.

cMERA is a unitary map taking the Gaussian reference state $\ket{R(M)}$ defined by Eqs.~\eqref{eq.vacuumNullifiers} and \eqref{eq.ReferenceState}, which is a product state with no spatial correlations,
to the approximate ground state $\ket{m^{(\Lambda)} }$  given by Eqs.~\eqref{eq.vacuumNullifiers} and \eqref{eq:alpha}.
One can view $\ket{R(M)}$ as the ground state of an \textit{ultra-local~Hamiltonian}
\begin{align}\label{eq.ultralocalH}
H_M = \int  d^{d}x : \left\{ \pi^{2} +  M^{2} \phi^{2} \right\}/2:\,,
\end{align}
where the $(\partial_{\vec{x}} \phi)^{2}$ term is omitted
and the mass $M$ is kept arbitrary.
Despite being a product state in real space, $\ket{R(M)}$ contains pairwise-entanglement in momentum space between momentum sectors $\vec{k}$ and $-\vec{k}$.

The cMERA circuit alters correlations between the $\vec{k}$ and $-\vec{k}$ modes from those corresponding to a constant and set by $M$ to the ones governed by $\alpha_{k}$ of Eq.~\eqref{eq:alpha} in a scale (i.e., $u$) dependent manner as follows
\small
\begin{align}
\label{eq.cMERA}
\ket{m^{(\Lambda)}} = \cP \, e^{-\frac{i}{2} \int_{-\infty}^{0} du \int_{k \leq \Lambda e^{u}\,} d^{d} k \, K(\vec{k}) \, \chi(u) } \, \ket{R(M)}\, .
\end{align}
\normalsize

The two mode squeezing operator $K(\vec{k})$ defined in Eq. \eqref{eq.defK} (dis)entangles the $\vec{k}$ and $- \vec{k}$ modes along the circuit. Energy minimization with respect to the free Hamiltonian $H_{m}$ as well as continuity of the transformation at the cutoff scale implies
\begin{equation}\label{chiAgain}
\hspace{-5pt}\chi(u) = \frac{1}{2} \frac{e^{2 u}}{e^{2 u} + {m^{2}}/{\Lambda^{2}}} ~~ \text{and} ~~ M = \omega_{\Lambda}\equiv\sqrt{\Lambda^2+m^2} .
\end{equation}
Comparing with Eq. \eqref{eq.defU} we see that the renormalization-group scale parameter $u$ running from the infrared at $u = -\infty$ to the ultraviolet at $u = 0$
plays the role of $\sigma\in\,[s_i,s_f]$ and $\int_{k \leq \Lambda e^{u}} d^{d} \vec{k} K(\vec{k}) \chi(u)/2 $ is equivalent to $G(s)$.
We also note that for every value of $u$, the action of the circuit on momenta larger than $\Lambda e^{u}$ is suppressed.
Since the operators $K(\vec{k})$ all commute, we may integrate over $u$  in Eq.~\eqref{eq.cMERA} to obtain
\begin{align}
\label{eq.cMERAcontract}
\ket{ m^{(\Lambda)} } = e^{- i \int_{k \leq \Lambda} d^{d} k \, K(\vec{k}) \, \log{\sqrt[4]{\frac{\omega_{\Lambda}}{\omega_{k}}}} } \, \ket{ R(\omega_{\Lambda}) }.
\end{align}
The above expression is the cMERA realization of Eq.~\eqref{eq.cMERAcontractGEN}.

\begin{figure}
\centering
\includegraphics[scale=0.51]{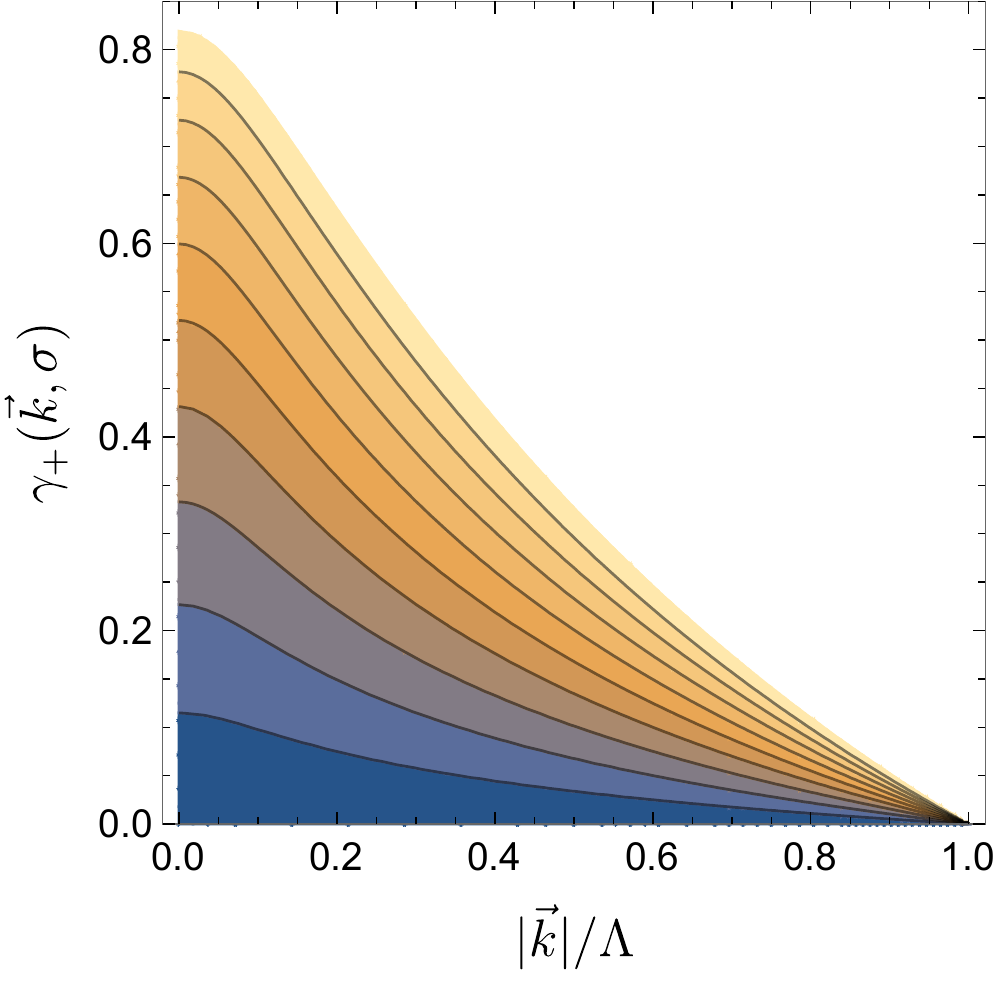}
~
\includegraphics[scale=0.45]{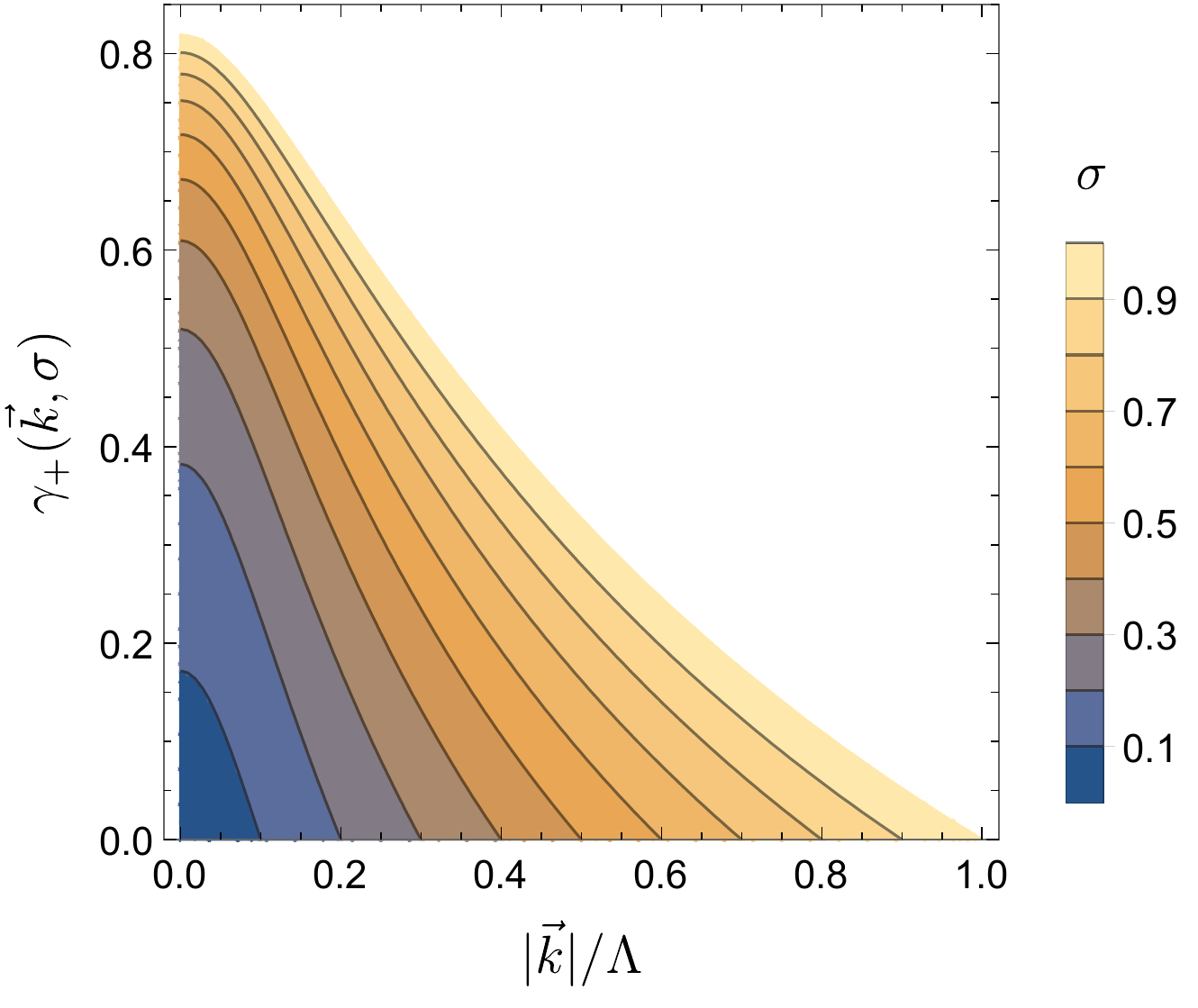}
\caption{Graphical description of the minimal circuit \eqref{minpathgamm} (left) and the cMERA circuit \eqref{cMERAgamma} (right). The plots present $\gamma_+$ as a function of $|\vec k|/\Lambda$ for different values of $\sigma$ represented by the different colored contours. We see that while the cMERA circuit only acts on momenta $|\vec k|/\Lambda<\sigma$, the minimal circuit alters $\gamma_+$ for all the different momenta at every step of the circuit. In this plot we have chosen $m/\Lambda=0.1$.}
\label{fig.circuit.compare}
\end{figure}

Evaluating the Fubini-Study metric distance according to Eq.~\eqref{eq.length} for the cMERA circuit yields
\begin{equation}
\ell_{\text{cMERA}}^{(2)} = \int_{-\infty}^{0} du \, \chi(u) \sqrt{\frac{\text{Vol}}{ 2} \int_{k < \Lambda e^u} d^d k}.
\end{equation}
We can evaluate this expression analytically and obtain
\begin{equation}
\frac{\Gamma \left(\frac{d}{2}+1\right)(\ell_{\text{cMERA}}^{(2)}) ^2}{2 \pi ^{d/2} \text{Vol} \,\Lambda ^d}   = \frac{\Lambda^4 \, _2F_1\left(1,\frac{d+4}{4};\frac{d+8}{4};-\frac{\Lambda ^2}{m^2}\right){}^2}{4 (d+4)^2 m^4}  ,
\end{equation}
where for the massless theory, the expression simplifies
\begin{equation}
\frac{\Gamma \left(\frac{d}{2}+1\right)}{2 \pi ^{d/2} \text{Vol} \, \Lambda ^d} \left( \ell_{\text{cMERA}}^{(2)} \right)^2 \biggr{|}_{m=0} = \frac{1}{4 d^2} \, .
\end{equation}
This is  equivalent to the complexity $\mathcal{C}^{(2)}$ of Eq.~\eqref{eq.C2explicit}.
Comparing to the relevant entry in Eq.~\eqref{eq.C4CFTsALL} (see \cite{SupMat}B below) we conclude  that the path generated by cMERA is a $\sqrt{2}$ longer than the minimal path associated with $K(\vec k)$, studied in the main text. Fig.~\ref{fig.circuit.compare} highlights the differences between these two circuits.

If we now disallow the different elementary generators in the cMERA circuit to act simultaneously and consider the equivalent of the $\mathcal{C}^{(1)}$ norm, we end up with
\begin{equation}
\ell_{\text{cMERA}}^{(1)} = \frac{\text{Vol}}{2} \int_{-\infty}^{0} du \, \chi(u)  \int_{k < \Lambda e^u} d^d k,
\end{equation}
which yields precisely the same circuit length as the one obtained in Eq.~\eqref{eq.C1NormNN} (see also Eq.~\eqref{eq.C4CFTsALL} in \cite{SupMat}B below). This is due to the fact that the $\mathcal{C}^{(1)}$ norm is invariant under independent reparametrizations of the circuits associated with the different pairs of momenta.

\section{B. Properties of the ground state complexities ${\cal C}^{(n)}$}
Here we analyze the properties of our complexity proposals \eqref{eq.C2explicit}-\eqref{eq.complexityproposal}.
In particular we focus on the structure of divergences. As we saw in the main text, the $\mathcal{C}^{(1)}$ norm results carry similarities to the results found using the holographic complexity proposals.

The leading divergence in the complexity measures ${\cal C}^{(n)}$
is proportional to $\mathrm{Vol}^{1/n}  \Lambda^{d/n}$ when $M=\Lambda$, and to $\mathrm{Vol}^{1/n}  \Lambda^{d/n}\log (M/\Lambda)$ when $M$ and $\Lambda$ are independent. The structure of subleading divergences depends on the interplay between $m$, $M$ and $\Lambda$ and we will analyze it in more detail for the $n = 1$ and $n = 2$ cases. For free CFTs ($m=0$) we obtain the following exact expressions
 \small
\begin{equation}\label{eq.C4CFTsALL}
\begin{split}
\hspace{-11 pt}
&\frac{
\Gamma\left( \frac{d}{2} + 1\right)}{\pi^{\frac{d}{2}} \Lambda^{d} \mathrm{Vol}}\cC^{(1)}_{CFT} =
\left\{
 \begin{array}{cl}
  \left| \log{\sqrt[4]{ \frac{M}{\Lambda} }} \right| + \frac{M^{d}}{2  d \Lambda^{d}} - \frac{1}{4  d},&  M < \Lambda
  \\
  \left| \log{\sqrt[4]{\frac{M}{\Lambda} }} \right| + \frac{1}{4 \, d}, & M \geq \Lambda
  \end{array}\right.
  \\
 \hspace{-11 pt}
&\frac{\Gamma\left( \frac{d}{2} + 1\right)
 }{2\pi^{\frac{d}{2}} \Lambda^{d} \mathrm{Vol}}
 \left( \cC^{(2)}_{CFT} \right)^{2} = \left|\log{\sqrt[4]{\frac{M}{\Lambda} }}\right|^2  + \frac{\log{\sqrt[4]{\frac{M}{\Lambda}}}}{2 d} + \frac{1}{8 d^2}.
\end{split}
\end{equation}
\normalsize
If we denote $M = e^{\gamma_{M}} \, \Lambda$, we see that the first terms in Eq.~\eqref{eq.C4CFTsALL} are indifferent to the sign of $\gamma_{M}$.
The remainder is smaller for a given value of $|\gamma_{M}|$ when $\gamma_{M}$ is negative and this  leads to a smaller complexity.
To understand this, note that for $M < \Lambda$ some $\vec{k}$-modes already begin with the right correlations with their $(-\vec{k})$-counterparts.
So our minimal unitary transformation can act much more mildly in the vicinity of this locus in momentum space which leads to a reduced complexity.
For the CFT, we see that the subleading divergences except for $\text{Vol} \, \Lambda^{d}$ and a constant vanish. This is similar to what happens in the CA proposal when curvature invariants over the relevant time slice vanish.

One can also consider the structure of divergences when the reference state is the CFT ($m=0$) vacuum and the target state is the vacuum of a free massive QFT.
This means that instead of considering $\ket{R(M)}$ as our reference state, we consider the vacuum of a free CFT $\ket{0}$ ($m=0$). The transformation between $\ket{0}$ and  $\ket{m^{(\Lambda)}}$ takes the form
\be
\label{eq.mfromCFT}
\ket{m^{(\Lambda)}} = e^{- i \int_{k \leq \Lambda} d^{d} k\, \bar r_k  K(\vec{k})
 } \ket{0};
~~~~ \bar r_k\equiv \log{\sqrt[4]{\frac{|\vec{k}|}{\omega_{k}}}}\, .
\ee
For the physically interesting case $\Lambda \gg m$, we see that the parameter of the transformation, $\bar r_k$, at large momenta approaches $0$ in accord with the physical intuition that mass becomes irrelevant in the UV. However, this does not mean that the complexity of the transformation does not diverge with the cut-off due to the growth of the number of momentum modes in the UV.
Adapting the $n = 1$ instance of our complexity proposal given by Eq.~\eqref{eq.C1NormNN} to the present case, i.e., replacing the $r_k \equiv \log{\sqrt[4]{M/\omega_{k}}}$ by $\bar r_k \equiv \log{\sqrt[4]{|\vec{k}|/\omega_{k}}}$,
we obtain that the complexity is now finite for $d = 1$, diverges logarithmically with the cut-off, $(m^2 \mathrm{Vol})\log{\frac{\Lambda}{m}}$, for $d = 2$ and in higher number of dimensions behaves as $(m^d \mathrm{Vol}) \left(\frac{\Lambda}{m}\right)^{d-2}$, for large values of the cutoff.
This behavior is subleading with respect to Eq.~\eqref{eq.C4CFTsALL} valid for starting with the unentangled reference state.
What we thus see is that taking the ground state of another QFT as a reference state
significantly lowers the complexity of the transformation needed to obtain $\ket{m^{(\Lambda)}}$, cf. Eq.~\eqref{eq.C4CFTsALL}, since some of the correlations have already been built, but still leads to UV-divergent results in space dimensions $d>1$.

\section{C. $\mathfrak{su}(1,1)$ manifold -- metric and geodesics}
Here, we provide the detailed derivation of the results presented in the main text for the ground state complexity using $\mathfrak{su}(1,1)$ generators.
In particular we study the general form of the manifold of states generated by the elements of the $\mathfrak{su}(1,1)$ algebras of Eq.~\eqref{eq.SU(1,1)a}, and the metric and geodesics induced on this manifold by the FS metric.

A path in the manifold of states will correspond to a path ordered exponential of the various $\mathfrak{su}(1,1)$ elements. For every point in the path such an exponent can always be regrouped as
\begin{equation}\label{unit112}
| \Psi (\sigma) \rangle = U(\sigma) | R (M) \rangle;
~~~~~U(\sigma)\equiv  e^{\int_{\Lambda} d^d k  \, g(\vec k,\sigma)} \, ,
\end{equation}
where
\begin{equation}\label{app:Unit1}
\begin{split}
 g(\vec k,\sigma) =\, &\alpha_{+}(\vec{k},\sigma) K_{+}(\vec k)
 \\
  & + \alpha_{-}(\vec k,\sigma) K_{-}(\vec k) + \omega(\vec k,\sigma) K_{0}(\vec k),
\end{split}
\end{equation}
and $\sigma$ is the path parameter from Eq.~\eqref{eq.defU}. This is due to the fact that the $\mathfrak{su}(1,1)$ generators form a closed algebra. The coefficients $\alpha_{\pm}(\vec k,\sigma)$ and $\omega(\vec k,\sigma)$ are not simply related to the original coefficients of the various generators in the path ordered trajectory \eqref{eq.defU} because of the noncommutative nature of generators.
Unitarity implies that $\alpha_{+}^{\ast}(\sigma) = - \alpha_{-}(\sigma)$ and $\omega^{\ast}(\sigma) =-\omega(\sigma)$.

The unitary \eqref{app:Unit1} can then be decomposed as follows (see e.g., \cite{Klimov2009} Appendix 11.3.3):
\begin{equation}\label{stateApp113}
\begin{split}
U(\sigma) = \, &e^{\int_\Lambda d^d k \gamma_{+} (\vec k,\sigma) K_{+}(\vec k)} \times
\\
&e^{\int_\Lambda d^d k \log \gamma_{0} (\vec k,\sigma) K_{0}(\vec k)} e^{\int_\Lambda d^d k \gamma_{-} (\vec k,\sigma) K_{-}(\vec k)} \, ,
\end{split}
\end{equation}
where the mapping between the coefficients is
\begin{align}\label{decomp1}
\begin{split}
  \gamma_{\pm} =\, & \frac{2 \, \alpha_{\pm} \sinh \Xi}{2 \, \Xi \cosh \Xi - \omega \sinh \Xi}\, ,
\\
 \gamma_{0} = \, &\left( \cosh \Xi - \frac{\omega}{2 \Xi} \sinh \Xi \right)^{-2} \,  ,
\\
 \Xi^2 \equiv \, &\frac{\omega^2}{4} - \alpha_{+} \alpha_{-} \, .
\end{split}
\end{align}
Note that unitarity implies that $|\gamma_{+}| < 1$, which fits nicely into the geometric picture describing the $\gamma_+$ manifold as a Poincar\'e {\it unit} disk.
We can now use the identities
\begin{equation}\label{KopOnR}
K_- | R (M) \rangle=0; ~~ K_0 | R (M) \rangle = \frac{\delta^{(d)}(0)}{4} | R (M) \rangle.
\end{equation}
to get rid of the information contained in the path parameters which only change our state by an overall phase.
This can be done by showing that the state in Eq.~\eqref{stateApp113} can be recast as
\begin{equation}\label{path21}
\begin{split}
 | \Psi (\sigma) \rangle = \, & \mathcal{N} e^{\int_\Lambda d^d k \gamma_{+} (\vec k,\sigma) K_{+}(\vec k)}| R (M) \rangle\, ,
 \\
 \mathcal{N} =\, & e^{\frac{\delta^{{}^{(d)}}(0)}{4} \int_\Lambda d^d k \log \gamma_0(\vec k, \sigma)}\, ,
\end{split}
\end{equation}
where $\mathcal{N}$ is a complex constant containing information about the overall normalization and phase.
The identity $|\gamma_0|=1-|\gamma_+|^2$ allows to demonstrate that up to an unphysical overall phase, the state depends only on $\gamma_+$.

One can then use the following set of adjoint conjugation identities to evaluate the Fubini-Study line element along the path \eqref{path21}
\begin{align}
\begin{split}
& e^{\int_\Lambda d^d k' \lambda(\vec k',\sigma) K_{+}(\vec k')} K_{0}(\vec{k}) e^{- \int_\Lambda d^d k'  \lambda(\vec k',\sigma) K_{+}(\vec k')}
\\
&\qquad =K_{0}(\vec k) - K_{+}(\vec k) \lambda(\vec{k},\sigma) \, ,  \\
&e^{\int_\Lambda d^d k' \lambda(\vec k',\sigma) K_{-}(\vec k')} K_{+}(\vec k) e^{-\int_\Lambda d^d k' \lambda(\vec k',\sigma) K_{-}(\vec k')}
\\
& \qquad
= K_{+}(\vec k) + 2 \lambda(\vec k,\sigma) K_{0}(\vec k) + \lambda(\vec k,\sigma)^2 K_{-}(\vec k)\, .
\end{split}
\end{align}
This leads to a remarkably simple form, resulting in the following expression for the complexity
\begin{equation}\label{EqAppTheMetric2}
\hspace{-1pt}\mathcal{C}= \min_{\gamma_+(\vec{k},\sigma)} \int_{s_i}^{s_f} d \sigma \sqrt{ \frac{\text{Vol}}{2} \int_{\Lambda} d^d \vec{k} \frac{ \gamma'_{+} (\vec{k}, \sigma)  \gamma'^{\ast}_{+} (\vec{k}, \sigma) }{(1- |\gamma_{+}(\vec{k}, \sigma)|^2  )^2} } ,
\end{equation}
where the prime denotes differentiation with respect to the path parameter $\sigma$.
We identify the line element of a manifold which consists of a direct product of hyperbolic unit discs, one for each pair of momenta with $\gamma_+$ for the different momenta playing the role of the complex coordinates on the discs.
The Poincar\'e unit disk is known to be the manifold associated with the coset $SU(1,1)/U(1)$ (see e.g., Refs. \cite{Perelomov:1971bd,Provost:1980nc,Perelomov:1986tf} and Ref. \cite{Molina-Vilaplana2015} in the context of cMERA). The $\mathfrak{su}(1,1)$ algebra whose generators are listed in Eq.~\eqref{eq.SU(1,1)a} generates isometries on the disk, with $K_0$ generating rotations around the origin and $K_1$ and $K_2$ generating pure translations along the imaginary and real axes respectively.
However note that the metric \eqref{LineElemFS:su11} couples the different speeds associated to the paths for different values of the momentum.

As explained in the main text, affinely parametrized geodesic paths on the product space correspond to affinely parametrized geodesics in each one of the spaces, where the relative speeds for the paths of the different momenta are dictated by the requirement that we reproduce the target state $\gamma_+ = \tanh(2 r_k)$ at the end of the path ($\sigma=s_f$). The geodesics on the Poincar\'e disk are well known and we identify the one connecting our reference state $\gamma_+=0$ to the target state as the solid line lying along the diameter in Fig.~\ref{PoincareDisk}.

The paths generated by $K(\vec k)$ in Eqs.~\eqref{eq.defU}, \eqref{eq.defG} and \eqref{eq.TheCircuit} and $B(\vec k,M)$ of Eq.~\eqref{Bapp} (see \cite{SupMat}D below) can be decomposed according to \eqref{unit112}-\eqref{decomp1} as follows
\begin{equation}\label{minpathgamm}
\begin{split}
K(\vec k): \qquad &\alpha_\pm(\vec k,\sigma) = \pm2 r_k\sigma, ~~ \omega(\vec k,\sigma)=0\,,
\\
& \gamma_+(\vec k,\sigma) = \tanh(2  r_k \sigma)\,,
\end{split}
\end{equation}
and
\begin{equation}
\begin{split}
B(\vec k,M): ~~~~~~~~&
\\
\alpha_\pm (\vec k,\sigma)  =& \frac{i \pi}{2} \sinh(2 r_k) \sigma\,,
\\
\omega(\vec k,\sigma) = & -i \pi \cosh(2 r_k) \sigma\,,
\\
\gamma_+(\vec k,\sigma) = &
\frac{i \sinh(2 r_k) \sin( \frac{\pi \sigma}{2})}{\cos(\frac{\pi \sigma}{2})+i \cosh(2 r_k)\sin(\sigma \frac{\pi}{2})}\,,
\end{split}
\end{equation}
where here we have taken $\sigma \in [0,1]$.
It is then possible to confirm that the path generated by $K(\vec k)$ is an affinely parametrized geodesic according to Eq.~\eqref{EqAppTheMetric2} while the one generated by $B(\vec k,M)$ is not a geodesic as it does not satisfy the Euler Lagrange equations derived from Eq.~\eqref{EqAppTheMetric2}.

For completeness we also specify the path corresponding to the cMERA circuit (see \cite{SupMat}A). After integrating the circuit (here it is possible since the generators commute) and decomposing according to Eq.~\eqref{decomp1} we obtain:
\begin{equation}\label{cMERAgamma}
\hspace{-7pt}\gamma_+(\vec k, \sigma) =
\tanh \left[ \frac{1}{4} \log \left[ \frac{m^2+\sigma^2 \Lambda^2}{k^2+m^2}\right]\right]\theta\left(\sigma-\frac{|k|}{\Lambda}\right),
\end{equation}
where we have redefined the cMERA parameter $\sigma\equiv e^u$. Fig.~2 is a comparison of the minimal path \eqref{minpathgamm} and the cMERA path \eqref{cMERAgamma}. It presents $\gamma_+$ as a function of $\vec k$ for different values of $\sigma$ and demonstrates in this way the progress of the circuit.

\section{D. Another constant generator}

What might be viewed as a deficiency of the transformation~\eqref{eq.cMERAcontractGEN} adopted from cMERA is that its generators~$K(\vec{k})$, see Eq.~\eqref{eq.defK}, do not have a spectrum bounded from below.
As such, they cannot be naturally interpreted as Hamiltonians of some fiducial physical system.
Here, we present an alternative constant generator  $B(\vec k,M)$ which induces constant period oscillations between the ground state and the reference state and does admit an interpretation as a lower bounded Hamiltonian and evaluate the associated path length.

The approximate ground state $|m^{(\Lambda)}\rangle$ of Eq.~\eqref{eq:alpha} can also be reached, up to an overall phase, starting from the reference state $\ket{R(M)}$ of Eq.~\eqref{eq.ReferenceState} by repeatedly applying the operator
\small
\begin{align}\label{Bapp}
B(\vec{k}, M) = -2\sinh(2r_k)[K_+ + K_-] + 4\cosh(2r_k)K_0.
\end{align}
\normalsize
Fig. \ref{PoincareDisk} contains a solid semicircle which illustrates the generated path.
Indeed, using the relevant decomposition formulas  Eqs.~\eqref{unit112}-\eqref{decomp1} (see e.g., \cite{Klimov2009} Appendix 11.3.3), one can establish that (cf. \eqref{eq.cMERAcontractGEN})
\begin{align}\label{BtransApp}
\ket{m^{(\Lambda)}} \simeq e^{- i \, \frac{\pi}{4} \int_{k \leq \Lambda} d^{d} k \, B\left(\vec{k}, \, M\right)} \ket{R(M)}\,,
\end{align}
where $\simeq$ indicates that the states are equal up to an irrelevant global phase.
Another way to obtain this transformation is to derive it using the properties of the Wigner distribution.
It is interesting to note that in contrast to Eq.~\eqref{eq.cMERAcontractGEN} the generators in Eq.~\eqref{Bapp} explicitly depend on both the reference and the target states.
However, the number of times each of them is applied in Eq.~\eqref{BtransApp} is fixed and equal to $\pi/4$.
Generalizing Eq.~\eqref{eq.complexityproposal} to the present case, we obtain for the different L$^{n}$ norms
\begin{align}
\begin{split}
\hspace{-2 pt}{\cal C}^{(n)}
= \frac{\pi}{4} \sqrt[n]{ \frac{\mathrm{Vol}}{2} \int_{k \leq \Lambda} d^{d} k \left| \sqrt{{M}/{\omega_{k}}} -\sqrt{{\omega_{k}}/{M}} \right|^n},
\end{split}
\end{align}
 which reads at leading order in $\Lambda$:
 \begin{align}
 \label{eq.C4CFTsnew}
 \begin{split}
& \frac{2^{2n+1} \Gamma\left( \frac{d}{2} + 1\right)
\left(\cC^{(n)} \right)^{n}}{\pi^{\frac{d}{2}+n} \, \Lambda^{d} \, \mathrm{Vol}}
\\
&~~~=
\begin{cases}
 \frac{2d }{2d+n}
 \left(\frac{\Lambda }{M}\right)^{n/2}+ \ldots,~~~~
  M \ll \Lambda,\\
 \frac{2d }{2d-n}  \left(\frac{M}{\Lambda }\right)^{n/2}
  + \ldots, ~~~~
  M \gg \Lambda,  ~2 d>n.
\end{cases}
\end{split}
\end{align}
One can clearly see that the leading divergences at large UV cut-off $\Lambda$ got now altered from $\Lambda^{\frac{d}{n}}$ to $\Lambda^{\frac{d }{n}\pm \frac{1}{2}}$ depending on the reference state scale $M$. Notice also that for $M \ll \Lambda$ ($M \gg \Lambda$), the number of gates $K(\vec{k})$ needed for the transformation from $\ket{R(M)}$ to $\ket{m^{(\Lambda)}}$ is smaller than the number of needed $B(M, \, \vec{k})$ gates.
This is in line with our predictions in the main text since $B(\vec k,M)$ deviates from the geodesic path generated by $K(\vec k)$.

To make this statement more precise, let us study the length of the minimal path constructed with the generator $B(\vec k,M)$ included inside the larger manifold spanned by the $\mathfrak{su}(1,1)$ generators of \eqref{eq.SU(1,1)a}
\begin{equation}\label{TheCircuitB2}
\ket{\Psi(\sigma)} \equiv e^{- i \,\sigma\, \frac{\pi}{4} \int_{k \leq \Lambda} d^{d} k \, B\left(\vec{k}, \, M\right)} \ket{R(M)}.
\end{equation}
This path can be shown to be minimal if only $B(\vec k, M)$ gates are allowed. According to the decomposition in Eqs.~\eqref{unit112}-\eqref{decomp1} we can show that this path corresponds to (see \cite{SupMat}C)
\begin{equation}
\begin{split}
\gamma_+(\vec k,\sigma) =
\frac{i \sinh(2 r_k) \sin(\sigma \frac{\pi}{2})}{\cos(\frac{\pi}{2}\sigma)+i \cosh(2 r_k)\sin(\sigma \frac{\pi}{2})}\,.
\end{split}
\end{equation}
Checking the Euler Lagrange equations explicitly for this path one concludes that it is not a geodesic. The path corresponding to $B(\vec k,M)$ is represented in Fig.~\ref{PoincareDisk} by a solid semicircle.
It would be interesting to explore if the operators $B(M, \, \vec{k})$ can lead to an alternative construction for a cMERA circuit.

\section{E. Fubini-Study metric derivation for a single $\mathfrak{su}(1,1)$}
Here, we present an alternative, simpler, derivation of Eq.~\eqref{EqAppTheMetric2} for a single  pair of momenta.
For this purpose, in the commutation relations of creation and annihilation operators, Dirac delta functions will be replaced by Kronecker deltas and integrals will be suppressed. We will restore them at the end of the calculation.  We will consider states of the form
\begin{equation}
\ket{\Psi(\sigma)} = \mathcal{N} e^{\gamma_{+}  K_{+}}| 0,0 \rangle\, ,
\end{equation}
where the state $|0,0\rangle$ contains no particle excitation in the $\vec{k}$ and $-\vec{k}$ modes according to the annihilation operators $b_{\vec k}$. In fact the state $\ket{R(M)}$ is a product of such states  $|0,0\rangle$ in the different momentum sectors. We can rewrite the state $\ket{\Psi(\sigma)}$ up to an overall phase as follows:
\begin{equation}
| \Psi (\sigma) \rangle \equiv \sqrt{1-|\gamma_+|^2} \sum_{n=0}^{\infty}\left( \gamma_{+} \right)^n |n, n \rangle \,,
\end{equation}
where we have fixed the constant by normalization and where $|n,n\rangle$ is the normalized state containing $n$ excitations with momentum $\vec{k}$ and $n$ with momentum $-\vec{k}$.
Small changes in $\gamma_+$ will result in the following change in the state $\ket{\Psi(\sigma)}$
\begin{equation}
\begin{split}
| \delta \Psi (\sigma) \rangle & =
\left(
\frac{\gamma_+ \delta \gamma_+^* + \gamma_+^* \delta \gamma_+}{\sqrt{1-|\gamma_+|^2}} \sum_{n=0}^{\infty}\left( \gamma_{+} \right)^n \right.
\\
&\left. +
\sqrt{1-|\gamma_+|^2} \sum_{n=0}^{\infty} n
\left( \gamma_{+} \right)^{n-1} \delta \gamma_+ \right)|n, n \rangle \,.
\end{split}
\end{equation}

Evaluating the Fubini-Study line element
\begin{equation}
  ds_{FS}^2  = \langle \delta \psi | \delta \nobracket \psi
  \rangle -\langle \delta \psi
  | \nobracket \psi \rangle \langle \psi | \delta \nobracket \psi
  \rangle \,,
\end{equation}
results in
\begin{equation}
 ds_{FS}^2 =  \frac{| \delta \gamma_+ |^2}{( 1 - |\gamma_+  |^2)^2}\, .
\end{equation}
Restoring the continuum structure, including the momentum integrals and delta functions one reaches  Eq.~\eqref{EqAppTheMetric2}. Note that $\text{Vol}/2$ that appears there is the volume of the space of pairs of momenta.

\section{F. Comparison with Ref. \cite{Jefferson:2017sdb}}
Ref.~\cite{Jefferson:2017sdb}, which appeared simultaneously with this article has some overlap with our results.
There, the authors use a lattice setup to study the complexity of the ground state of a free scalar field theory.
An important difference between our approach and the one used in Ref.~\cite{Jefferson:2017sdb} is the nature of the distance metric minimized to obtain complexity.
While in our approach, we minimize the Fubini-Study distance over states restricted to the manifold carved by allowed generators, in Ref.~\cite{Jefferson:2017sdb} the distance measure is over unitaries in the spirit of \cite{Nielsen2006}.
In addition, the set of gates we consider here is different from the one in Ref. \cite{Jefferson:2017sdb}. However, the two sets contain the gates $K(\vec k)$ of Eq.~\eqref{eq.defK}, which turns out to be the one which in both approaches is used to construct the minimal length circuit.  We compare below the various components of the construction, i.e., the gates, the distance function, the regularization method and the final result.

{\bf Distance measures:} We use the Fubini-Study metric over states to measure the length of our path while Ref.~\cite{Jefferson:2017sdb} considers metrics in Finsler geometry over unitary circuits inspired by Nielsen's approach, see Ref. \cite{Nielsen2006}.
Being based on the projective nature of the Hilbert space, the FS metric does not account for generators acting trivially on the state, namely generators which only modify the state up to an overall phase, while this is not in general the case for a unitary based approach.
For the gates $K(\vec k)$ of Eq.~\eqref{eq.defK} the two approaches give the same result up to an overall numerical factor. It would be interesting however to explore to what extent they yield similar notions of complexity when studying complexity of pure Gaussian states with a larger set of gates. For instance it would be interesting to understand if the geometry remains hyperbolic when considering the full set of $\mathfrak{su}(1,1)$ generators using the approach of Ref. \cite{Jefferson:2017sdb}.
For this it is important to understand how to extend the Nielsen construction and assign a ``weight'' to gates constructed from the different generators of $\mathfrak{su}(1,1)$.

{\bf Generator sets:}
Both works consider quadratic generators in order to build the allowed unitary transformation on the state. Gates constructed from these generators transform Gaussian states  among themselves which allows to restrict the analysis to Gaussian states.
Ref. \cite{Jefferson:2017sdb} minimizes over all gates constructed by exponentiating  bilinear generators of the form $\phi(x)\pi(y)$.
In fact, the authors do this on the lattice for generators $G_{ab}=x_a p_b + p_b x_a$ where $a$ and $b$ enumerate the lattice sites which form a  $GL(N,{\rm I\!R})$ group structure for a one dimensional lattice with $N$ sites.
In the absence of \textit{penalty factors} (Penalty factors correspond to modifying the cost function associated to a generator in order to favor/penalize certain generators; this could for example be used to penalize long range generators with respect to local ones), Ref. \cite{Jefferson:2017sdb} finds that an optimal circuit admits a {\it  normal mode} decomposition and thus requires only momentum preserving generators of the form $\phi(k) \pi(-k)+\pi(k)\phi(-k)$ for its construction.
On the lattice these are given by $G_{k}=\tilde x_k \tilde p_{-k}+\tilde p_{k}\tilde x_{-k}$, where $\tilde x_k \equiv \frac{1}{\sqrt{N}} \sum_{a=0}^{N-1} \exp\left(-\frac{2 \pi i k \, a}{N}\right)x_a$.
Inspired by cMERA, our starting point was to consider (a larger set of) momentum preserving generators.
In this sense, our work is complementary to that of Ref. \cite{Jefferson:2017sdb}.
We minimized over gates constructed from the following set of generators
\begin{align*}
\begin{split}
&G_{1,k} = \phi(\vec k) \pi(-\vec k) + \pi(\vec k) \phi(-\vec k);
\\
&G_{2,k} = \phi(\vec k) \phi(-\vec k);
~
G_{3,k} = \pi(\vec k) \pi(-\vec k);
\end{split}
\end{align*}
which form $\mathfrak{su}(1,1)$ subalgebras of quadratic generators conserving momentum.
These are bi-local in real space and contain the generator $G_{1,k}$ (equivalently, $K(\vec k)$ of Eq.~\eqref{eq.defK}) of cMERA circuits.
In both works it was found that the preferred motion is in the direction of $G_{1,k}$ which lies in the intersection of the two sets of generators.
It would be  interesting to minimize the complexity using an algebra of generators which accounts for both our gates and those of Ref.~\cite{Jefferson:2017sdb} and check whether the minimal circuit is still generated by $G_{1,k}$.

{\bf Regulating divergences:} Our method of regulating divergences was that our circuit only reproduced the ground state faithfully up to the cutoff momentum. This allowed us to obtain finite results. Ref.~\cite{Jefferson:2017sdb} uses a lattice regularization in order to obtain finite results and the regulator is set by the lattice spacing.

{\bf Final result:} The finite result is very similar, the main difference (except for an overall factor) is due to the different regularization schemes.
Recall that we obtained in Eq.~\eqref{eq.C2explicit}
\begin{align}
{\cal C}^{(2)} = \frac{1}{2}\sqrt{ \frac{\mathrm{Vol} }{2}  \int_{k \leq \Lambda} d^{d} k\,   \left(\log\frac{\omega_k}{M}\right)^{2}}.
\end{align}
This is to be compared to the result in Eqs.~(4.32)-(4.33) of Ref.~\cite{Jefferson:2017sdb} which reads
\begin{align}
\mathcal{C}^{(2)}=\frac{1}{2}\sqrt{\sum_{k_i=0}^{N-1}\left(\log\frac{\tilde\omega_{\vec k}}{\omega_0}\right)^2},
\end{align}
with the normal mode frequencies given by
\begin{align}
\tilde\omega_{\vec k}^2 = m^2 +\frac{4}{\delta^2} \sum_{i=1}^{d-1}\sin^2\frac{\pi k_i}{N}.
\end{align}
We see that $\delta$ the lattice spacing in Ref.~\cite{Jefferson:2017sdb} and $\Lambda$ the momentum cutoff in our work play a similar role. We also see that the reference state scale $M$ is essentially the same as the reference state parameter $\omega_0$ in Ref. \cite{Jefferson:2017sdb}. The leading divergence in $C^{(1)}$ in Eq.~\eqref{eq.LnNormsDivB} (see also \eqref{eq.C4CFTsALL}) can be compared to the one obtained in Eqs.~(E.11)-(E.12) of Ref.~\cite{Jefferson:2017sdb} and is found to be the same up to an overall numerical factor.

\end{document}